\pdfoutput=1
\documentclass[journal=accacs,manuscript=article]{achemso}
\usepackage{xcolor}
\usepackage{tabularx}
\usepackage{soul}
\usepackage{threeparttablex} % table with notes
\usepackage{longtable} % table
\usepackage{rotating} % rotating tables and figures
\usepackage{amsmath}
\usepackage{comment}
\usepackage{gensymb}
\usepackage[version=4]{mhchem}

\author{Jonathan W. P. Zajac}
\affiliation[UMN]{Department of Chemistry, University of Minnesota, Minneapolis, MN 55455, USA}
\alsoaffiliation[CTC]{Chemical Theory Center, University of Minnesota, Minneapolis, MN 55455, USA}

\author{Praveen Muralikrishnan}
\affiliation[CEMS]{Department of Chemical Engineering and Materials Science, University of Minnesota, Minneapolis, MN 55455, USA}
\alsoaffiliation[CTC]{Chemical Theory Center, University of Minnesota, Minneapolis, MN 55455, USA}

\author{Caryn L. Heldt}
\affiliation[MTU]
{Department of Chemical Engineering, Michigan Technological University, Houghton, MI 49931, USA}

\author{Sarah L. Perry}
\affiliation[UMA]
{Department of Chemical Engineering, University of Massachusetts Amherst, MA 01003, USA}

\author{Sapna Sarupria}
\email{sarupria@umn.edu}
\affiliation[UMN]{Department of Chemistry, University of Minnesota, Minneapolis, MN 55455, USA}
\alsoaffiliation[CTC]{Chemical Theory Center, University of Minnesota, Minneapolis, MN 55455, USA}

\date{\today}

\title{Impact of Co-Excipient Selection on Hydrophobic Polymer Folding: Insights for Optimal Formulation Design}

\begin{document}
\doublespacing
\clearpage

\begin{abstract}
The stabilization of liquid biological products is a complex task that depends on the chemical composition of both the active ingredient and any excipients in solution. Frequently, a large number of unique excipients are required to stabilize biologics, though it is not well-known how these excipients interact with one another. To probe these excipient-excipient interactions, we performed molecular dynamics simulations of arginine -- a widely used excipient with unique properties -- in solution either alone or with equimolar lysine or glutamate. We studied the effects of these mixtures on a hydrophobic polymer model to isolate excipient mechanisms on hydrophobic interactions, relevant to both protein folding and biomolecular self-assembly. We observed that arginine is the most effective single excipient in stabilizing hydrophobic polymer collapse, and its effectiveness can be augmented by lysine or glutamate addition. We utilized a decomposition of the potential of mean force to identify that the key source of arginine-lysine and arginine-glutamate synergy on polymer collapse is a reduction in attractive polymer-excipient direct interactions. Further, we applied principles from network theory to characterize the local solvent network that embeds the hydrophobic polymer. Through this approach, we found that arginine enables a more highly connected and stable network than in pure water, lysine, or glutamate solutions. Importantly, these network properties are preserved when lysine or glutamate are added to arginine solutions. Overall, we highlight the importance of identifying key molecular consequences of co-excipient selection, aiding in the establishment of rational formulation design rules.
\end{abstract}

\clearpage
\section{Introduction}
Biologics are complex pharmaceuticals including proteins, vaccines, plasma products, gene therapy vectors, and biological tissues.\cite{strassburg_global_1982, zhang_first_2018} The stability of biologics is primarily concerned with physical denaturation related to protein unfolding and aggregation, among other factors.\cite{privalov_cold_1990, heremans_high_1982, sarupria_studying_2010,  kauzmann_factors_1959, frank_free_1945, christensen_denaturation_1952, hummer_pressure_1998, ghosh_molecular_2001, le_basle_physicochemical_2020, authelin_freezing_2020, crommelin_formulation_2024, clarkson_conformational_2016, huynh-ba_stability_2009} These formulations are especially susceptible to degradation when exposed to environmental stressors such as elevated temperatures, often necessitating refrigerated storage.\cite{world_health_organization_temperature_2006, hanson_is_2017, yu_grand_2021, crommelin_addressing_2021} The logistics workflow known as the cold chain maintains biologics at cold temperatures from manufacturing, to long term storage, to distribution and outreach programs. However, equipment breakdowns and improper training at any point along the cold chain can disrupt entire batches of biologics, adverself affecting the accessibility of these products.\cite{thielmann_visual_2019, thielmann_improving_2020, crommelin_addressing_2021} 

A promising strategy to relieve the cold chain burden involves designing biological formulations with additives known as excipients.\cite{kamerzell_proteinexcipient_2011, arakawa_biotechnology_2007, bongioanni_amino_2022, jeong_analytical_2012, wang_protein_2005, patro_protein_2002, ionova_biologic_2020} Excipients are often small molecules such as amino acids or sugars. Studies of excipient effects on the stability of biologics are generally high-throughput and empirical in nature, often lacking the resolution required to understand the molecular mechanism at play. Additionally, excipients are, on average, deployed in combinations of four or more unique molecules.\cite{ionova_biologic_2020} This results in a vast design space and a generally poor understanding of individual excipient properties and combinatorial effects. We hypothesize that understanding the molecular details of excipient mechanisms will aid in predicting optimal excipient selections for novel formulations.

Motivated to understand the molecular-level effects of excipient combinations in biologics stability, we utilized molecular dynamics (MD) simulations of a hydrophobic polymer as a model for protein folding/unfolding. Hydrophobic polymer models are effective in decoupling hydrophobic interactions from other forces involved in protein folding (e.g., electrostatic, van der Waals, and hydrogen bonding interactions),\cite{ten_wolde_drying-induced_2002, zangi_ureas_2009, nayar_cosolvent_2018, mondal_when_2013, athawale_enthalpyentropy_2008, jamadagni_how_2009, Zajac2024arXiv} a daunting task experimentally.  By isolating hydrophobic interactions, fundamental insights into excipient mechanisms on protein stability can be obtained.\cite{timasheff_control_1993, ghosh_salt-induced_2005, athawale_osmolyte_2005, athawale_enthalpyentropy_2008, zangi_ureas_2009, canchi_cosolvent_2013, van_der_vegt_hydrophobic_2017, Zajac2024arXiv} For example, the denaturant urea weakens hydrophobic interactions,\cite{wallqvist_hydrophobic_1998, ikeguchi_molecular_2001, ghosh_salt-induced_2005, van_der_vegt_enthalpyentropy_2006, lee_does_2006, athawale_enthalpyentropy_2008, zangi_ureas_2009, shpiruk_effect_2013} while the stabilizing osmolyte trimethylamine N-oxide (TMAO) negligibly affects or strengthens these effects.\cite{athawale_osmolyte_2005, paul_hydrophobic_2008, macdonald_effects_2013, ganguly_hydrophobic_2016, su_effects_2018, folberth_small--large_2022}

As a starting point towards establishing excipient design rules, we turned our attention to a widely used excipient, arginine (Arg). Arg is a versatile excipient with a wide range of reported effects on the stability of biologics. Arg is frequently used in protein and vaccine storage, purification techniques, and as an aggregation reducer.\cite{tsumoto_role_2004, arakawa_biotechnology_2007, tsumoto_review_2005, startzel_arginine_2018, hamborsky_epidemiology_2015, mistilis_long-term_2017} However, in some contexts, Arg has been found to denature proteins,\cite{xie_guanidine_2004, anumalla_counteracting_2019, arakawa_effects_2018} accelerate aggregation,\cite{smirnova_l-arginine_2013, shah_effects_2011, eronina_dual_2014} and inactivate viruses.\cite{meingast_arginineenveloped_2020, meingast_physiochemical_2021, Zajac2024arXiv} In situations where Arg is denaturing, addition of charged co-excipients has been observed to reverse the denaturing properties of Arg.\cite{anumalla_counteracting_2019} In other contexts, synergy has been observed between Arg and glutamate (Glu) in the solubilization of model proteins.\cite{shukla_understanding_2011} Recently, we proposed that the multi-faceted effects of Arg arise from its positioning at the edge of a mechanistic flip between indirect- and direct-dominated mechanisms on hydrophobic interactions.\cite{Zajac2024arXiv} Due to its placement at this edge, we aimed to understand whether subtle changes in the formulation environment alters the stabilizing properties of arginine. To this end, we investigated hydrophobic polymer folding in lysine (Lys), Glu, and Arg solutions, as well as equimolar formulations of Arg/Lys, Arg/Glu, and Lys/Glu. We discovered that, adding Lys or Glu to Arg solutions enhances hydrophobic polymer stability -- underscoring the importance of co-excipient selection of biologics formulation design.

\section{Methods}
\subsection{System Setup and Molecular Dynamics Simulations}
To probe the effects of Arg, Lys, Glu, and binary excipient mixtures on hydrophobic interactions, we completed MD simulations of a hydrophobic polymer in excipient solutions at various concentrations. Replica exchange umbrella sampling (REUS)\cite{Sugita_REUS_2000} simulations were utilized to calculate the potential of mean force (PMF) of hydrophobic polymer folding in different excipient solutions. The hydrophobic polymer was modeled as a linear coarse-grained chain with 26 monomers. Each monomer represents a methane-like unit, with Lennard-Jones parameters $\sigma= 0.373$ $\mathrm{nm}$ and $\epsilon= 0.5856$ $\mathrm{kJ/mol}$.\cite{athawale_osmolyte_2005}  The polymer-water $\epsilon$ parameter was modified to achieve an approximately even distribution of folded and unfolded polymer states in pure water.\cite{Zajac2024arXiv} Box dimensions were defined such that 1.5 nm of space separated the fully elongated polymer from the nearest box edge. All systems were solvated with TIP4P/2005 water.\cite{TIP4P/2005} The salt forms of all excipients (Arg/Cl, Lys/Cl, Glu/Na) under study were added to the box until the desired concentration was reached (Table S1). The CHARMM22 force field was used to describe excipient molecules and ions.\cite{Brooks1983CHARMMAP, brooks_charmm_2009} With this protocol, we generated systems comprised of 0.25 M, 0.5 M, and 1.0 M Arg, Lys, Glu, Arg/Lys, Arg/Glu, and Lys/Glu. In binary excipient solutions, equimolar concentrations were used. In excipient solutions with no polymer present, the same box size as in the polymer systems was used.

All simulations were initially subject to energy minimization using the steepest descent algorithm. 1 ns NVT equilibration simulations were carried out at 300 K, followed by 1 ns NPT equilibration simulations at 300 K and 1 atm. During equilibration, temperature was controlled according to the V-rescale thermostat\cite{bussi_donadio_parrinello_2007}, while pressure was controlled via the Berendsen barostat.\cite{berendsen_postma_van_gunsteren_dinola_haak_1984}  Following equilibration, NPT production runs were completed using the Nos\'e-Hoover thermostat ($\mathrm{\tau_{T}}$ = 5 $\mathrm{ps}$) \cite{evans_holian_1985} and Parrinello-Rahman barostat ($\mathrm{\tau_P}$ = 25 $\mathrm{ps}$).\cite{parrinello_rahman_1981} Production runs were 20 ns long for excipient/water systems, and between 50-250 ns per window for excipient/polymer/water REUS simulations (Table S1). In all simulations, the Particle Mesh Ewald (PME) algorithm was used for electrostatic interactions with a cut-off of 1 $\mathrm{nm}$. A reciprocal grid of 42 x 42 x 42 cells was used with $\mathrm{4^{th}}$ order B-spline interpolation. A single cut-off of 1 $\mathrm{nm}$ was used for Van der Waals interactions. The neighbor search was performed every 10 steps. Lorentz-Berthelot mixing rules\cite{lorentz_1881, berthelot1898melange} were used to calculate non-bonded interactions between different atom types.  All simulations were run in GROMACS 2021.4.\cite{Lindahl_Spoel_2021}
 
\subsection{Replica Exchange Umbrella Sampling}
REUS\cite{Sugita_REUS_2000} simulations were completed to sample the hydrophobic polymer conformational landscape in excipient solutions.  REUS simulations were completed using GROMACS 2021.4\cite{Lindahl_Spoel_2021} with the PLUMED 2.8.0 \cite{PLUMED_Consortium_2019,Tribello_Bussi_2014} patch applied. The radius of gyration ($R_{g}$) of the hydrophobic polymer was used as a reaction coordinate, placing 12 umbrella potential centers evenly between $R_{g}$ = 0.3 and $R_{g}$ = 0.9 nm. A force constant of $K$ = 5000 kJ/mol/nm$^2$ was used in all windows, with the exception of the window centered at $R_g$ = 0.45, which used $K$ = 1000 kJ/mol/nm$^2$.\cite{Zajac2024arXiv}

The potential of mean force (PMF) of polymer folding/unfolding was calculated as ${W(R_g) = -k_{B}T ln(P(R_g))}$. Biased probability distributions were reweighted according to the Weighted Histogram Analysis Method (WHAM).\cite{zhu_hummer_2011} The free energy of polymer unfolding ($\Delta G_u$) was calculated according to:
\begin{equation}\label{unfolding_free_energy}
     \exp \left(\frac{\Delta G_{\text{u}}}{k_BT}\right)=\frac{\int_{R_{g,cut }}^{R_{g,\max}} \exp\left(\frac{-W\left(R_g\right)}{k_B T}\right) dR_g}{\int_{R_{g,\min}}^{R_{g,cut }} \exp\left(\frac{-W\left(R_g\right)}{k_B T}\right) dR_g}
\end{equation}
where $R_{g,cut}$ was determined as the point between the folded and unfolded states where $
\frac{\partial W(R_{g})}{\partial R_{g}} = 0$. 

Following the methods outlined by several others,\cite{athawale_effects_2007, godawat_unfolding_2010, van_der_vegt_hydrophobic_2017, dasetty_advancing_2021, Zajac2024arXiv} the PMF was decomposed as
\begin{equation}
           W(R_g) = W_{vac}(R_g)+ W_{cav}(R_g ) + E_{pw}(R_g) + E_{pa} (R_g)+ E_{pc}(R_g)
\label{eqn:decomp}
\end{equation}
$W_{vac} (R_g)$ captures intrapolymer degrees of freedom and was obtained from independent REUS simulations of the polymer in vacuum. $E_{pw} (R_g)$, $E_{pa} (R_g)$, and $E_{pc} (R_g)$ are average polymer-water, polymer-additive, and polymer-counterion interaction energies, respectively. The remaining term is $W_{cav} (R_g)$, which is the cavitation component and quantifies the energetic cost of forming a polymer-sized cavity in solution.

\subsection{Preferential Interaction Coefficients}
Distribution of water and excipient molecules with respect to any solute can be quantified via the preferential interaction coefficient, $\Gamma_{PA}$.\cite{Scatchard_1946, Casassa_1964, Schellman_1987} This parameter is calculated in simulations using the two-domain formula:\cite{inoue_timasheff_1972, record_anderson_1995, shukla_molecular_2009}
\begin{equation}
    \Gamma_{PA}=\left\langle N_A^{\text {local }}-\left(\frac{N_A^{\text {bulk }}}{N_W^{\text {bulk }}}\right) N_W^{\text {local }}\right\rangle
\end{equation}
\noindent
where $P$ denotes the polymer, $A$ represents an additive species (Arg, Lys, Glu, Na$^{+}$, or Cl$^{-}$), and $W$ denotes water. $N$ represents the number of molecules of a given species, while angular brackets denote an ensemble average. The local and bulk domain was separated by a cutoff distance $R_{cut}$ from the polymer. $\Gamma_{PA}$ gives a measure of the relative accumulation or depletion of an additive in the local domain of the hydrophobic polymer, with $\Gamma_{PA} > 0$ indicating relative accumulation (preferential interaction) and $\Gamma_{PA} < 0$ indicating relative depletion (preferential exclusion).

Wyman-Tanford theory relates any equilibrium process and preferential interaction as:\cite{wyman_linked_1964, timasheff_protein-solvent_2002, shukla_molecular_2011}

\begin{equation}
    -\left(\frac{\partial\Delta G_{u}}{\partial\mu_{A}}\right) = \Gamma_{PA}^{u} - \Gamma_{PA}^{f}
\end{equation}
\noindent Here, we use this relationship to connect preferential interactions in the unfolded ($\Gamma_{PA}^{u}$) and folded ($\Gamma_{PA}^{f}$) ensembles As a result of this relationship, denaturants are expected to have a greater preferential interaction coefficient in the unfolded ensemble, while stabilizing osmolytes have a greater preferential interaction coefficient in the folded ensemble.\cite{canchi_cosolvent_2013, mondal_when_2013, mondal_how_2015, mukherjee_unifying_2020}

\subsection{Arginine Clustering}
Several studies have identified the importance of Arg clustering in the variable effects of the excipient.         \cite{das_inhibition_2007, schneider_investigation_2009, shukla_interaction_2010, schneider_arginine_2011, vagenende_protein-associated_2013, santra_analyzing_2021, santra_influence_2022, meingast_physiochemical_2021} As a free molecule, Arg forms self-associated clusters via three primary interactions: (i) backbone-backbone (COO$^{-}$--NH$_{3}^{+}$), (ii) backbone-sidechain (Gdm$^{+}$--COO$^{-}$), and (iii) sidechain-sidechain (Gdm$^{+}$--Gdm$^{+}$). To quantify the extent of Arg cluster formation, we applied the following geometric criteria for the interactions defined above between pairs of molecules $i$ and $j$, where $i \neq j$: (i) at least one COO$^{-}$ oxygen from $i$ within 2.0 \AA~of an NH$_{3}^{+}$ hydrogen from $j$, (ii) at least one COO$^{-}$ oxygen from $i$ within 2.0 \AA~of a Gdm$^{+}$ hydrogen from $j$, and (iii) at least one Gdm$^{+}$ carbon from $i$ within 4.0 \AA~of a Gdm$^{+}$ carbon from $j$.

For binary excipient solutions, criteria (i) and (ii) may be met via the sidechains of Glu and Lys, which introduce additional COO$^{-}$ and NH$_{3}^{+}$ groups into the system, respectively. Criterion (iii) may only be achieved via two interacting Arg molecules. For every excipient molecule $i$ in solution, we iteratively searched over every other excipient molecule $j$. Molecules found to match the criteria outlined above were used to construct individual graphs, $g_{i}$, with a central node positioned on molecule $i$ and edges connecting $i$ to all interacting residues $j$. NetworkX\cite{osti_960616} was used to merge any individual graphs $g_{i}$ with shared edges into clusters, $c_{i}$. The largest cluster size is reported as the maximum value of elements within any of the $c_{i}$ constructed clusters.

To characterize excipient clusters according to interaction types, we computed an interaction efficiency metric, $\eta$, according to:
\begin{equation}
\begin{split}
           \eta = n_{k} / N_{k}
\end{split}
\label{eqn:cluster}
\end{equation}
\noindent where $n_{k}$ is the number of contacts observed that match criteria $k$, while $N_{k}$ is the total number of all excipient molecules that can participate in criteria $k$. For criteria (i) and (ii), $N_{k}$ denotes the total number of excipient molecules in solution, while for (iii), $N_{k}$ denotes the total number of arginine molecules, as only Arg molecules can satisfy this criterion.

\subsection{Network Analysis}
Several techniques from network theory were applied to quantify the solvent structure of the local polymer domain. Graphs, $G(t)$, were constructed for a configuration at time $t$ using NetworkX.\cite{osti_960616} Nodes were defined as any solvent molecule center-of-mass within 0.7 nm of the hydrophobic polymer. An edge was constructed between nodes if a pair of heavy atoms $i$ and $j$, belonging to nodes representing residues $I$ and $J$, were within 0.35 nm of each other. Configurations were taken every 100 ps from the completed REUS trajectories.

We used the Wasserman and Faust improved formula to calculate closeness centrality for all nodes in the graph:\cite{wasserman_social_1994, freeman_centrality_1978}

\begin{equation}
\begin{split}
           C_{c}(u) = \frac{n-1}{N-1} \left(\frac{n-1}{\sum_{v=1}^{n-1}d(u,v)}\right)
\end{split}
\label{eqn:closeness}
\end{equation}

\noindent where $n$ is the number of reachable nodes from node $u$, and $N$ is the total number of nodes in the graph, $G(t)$. $d(u,v)$ is the shortest distance between node $u$ and reachable node $v$. A reachable node refers to any node that is accessible to node $u$ through a continuous sequence of adjacent nodes. Betweenness centrality was measured as:\cite{freeman_centrality_1978, brandes_faster_2001, brandes_variants_2008}

\begin{equation}
\begin{split}
           C_{b}(u) = \sum_{s,t} \frac{np(s,t|u)}{np(s,t)}
\end{split}
\label{eqn:betweenness}
\end{equation}

\noindent where $np(s,t)$ is the number of shortest paths between nodes $s$ and $t$ through the connected network, and $np(s,t|u)$ is the number of such paths that pass through node $u$. For each centrality measurement, the average value in each graph $G(t)$ was reported and plotted in a distribution that captures all frames of the trajectory. Further, by assigning nodes as either belonging to excipient molecules or water molecules, we decomposed these quantities into water centrality ($C^{wat}$) and excipient centrality ($C^{exc}$) measurements.

We measured graph stability by computing a fragmentation threshold, $f$. In this approach, we began with all complete graphs $G(t)$ for which the number of independent graphs was equal to 1. Iteratively, we randomly removed individual nodes and any associated edges from the graph. At each step, the number of disconnected graphs was computed, and the point at which this value changed from 1 to 2 was recorded as the fragmentation threshold. In practice, this value is reported as the fraction of nodes removed, $f = u/U$, where $u$ is the number removed and $U$ is the total number of nodes in the graph.

\subsection{Hydration Shell Dynamics}
The rotational dynamics of water was measured by computing the characteristic reorientation time of the water dipole, $\mu$.\cite{liu_structural_2021, diaz_effect_2023, santra_influence_2022} This dipole was taken as the vector connecting the oxygen atom and the center of the two hydrogen atoms of water molecule $w_{i}$. The time evolution of this vector was monitored by computing the time correlation function:

\begin{equation}
\begin{split}
           C_{\mu}(t) = \frac{\langle\mu_{i}(0)\cdot\mu_{i}(t)\rangle}{\langle\mu_{i}(0)\cdot\mu_{i}(0)\rangle}
\end{split}
\label{eqn:reorientation}
\end{equation}

\noindent where $\mu_{i}(t)$ is the dipole vector of the $i$th water molecule at time $t$. Water molecules were considered for analysis according to the following protocol (Scheme S1): (i) the Cartesian coordinates of the oxygen water falls within $r$ to $r+dr$ at $t=0$, (ii) if a water molecule moves into a buffer region, spanning $r+dr$ to $r+dr+b$, its position is flagged and tracked over time, and (iii) a water molecule is removed from consideration if the tracked molecule exits the buffer region without re-entry into the $r$ to $r+dr$ shell, or persists within the buffer region for at least 2 ps. In the above criteria, $r$ denotes the minimum distance from the hydrophobic polymer, $dr$ is the width of the hydration layer under consideration, and $b$ is the width of the buffer region.

To ensure accurate sampling, we used three distinct starting configurations per system for water dynamics analysis. Polymer configurations were clustered using HDBSCAN\cite{mcinnes_hdbscan_2017}, where clusters representing the free energy minima of the hydrophobic polymer PMF were identified (Fig. S1, Fig. S2) Configurations with the highest cluster membership probability were selected as starting configurations. From these starting points, 300 ps NPT production runs were completed, saving configurations every 0.1 ps.

\section{Results and Discussion}
The goal of this study is to elucidate the mechanisms underpinning the use of amino acids in the stabilization of biologics. To this end, we examined the effect of both single solutions of the amino acids Arg, Lys, Glu, and binary mixtures of these species on the stability of a hydrophobic polymer model.

\subsection{Hydrophobic Polymer Collapse is Favorable in Excipient Solutions}
\begin{figure*}[!hb]
   \includegraphics[width=1.0\textwidth]{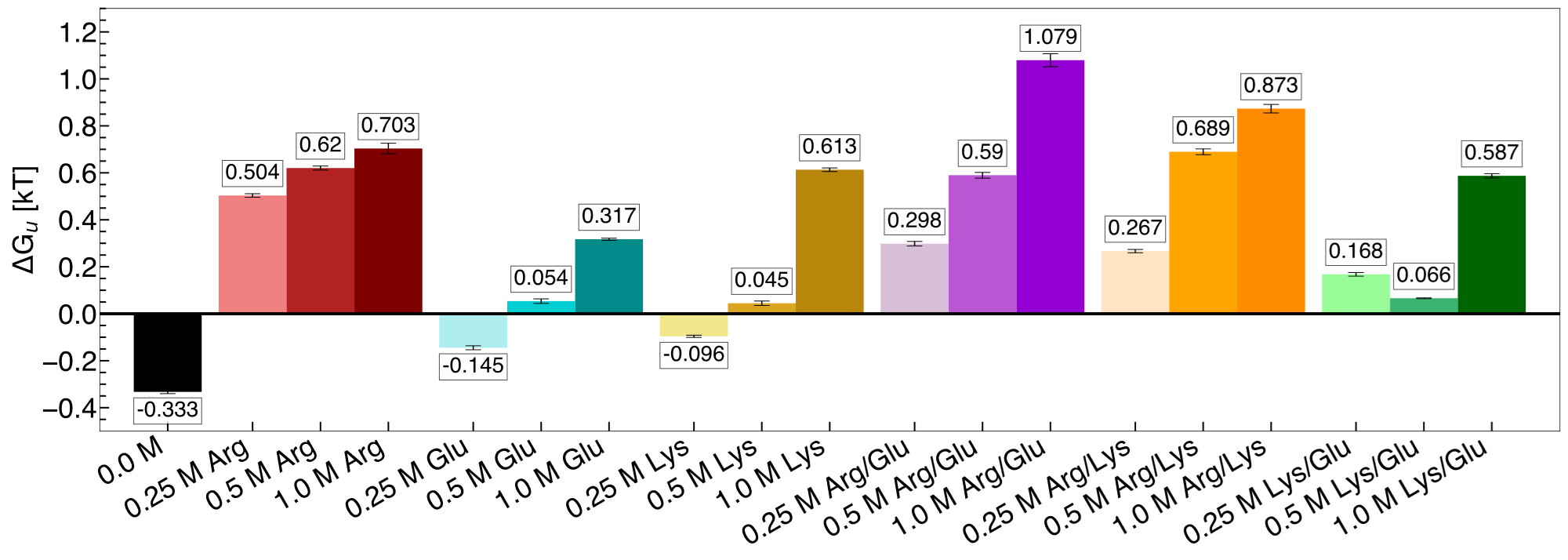}
    \caption{Free energy of hydrophobic polymer unfolding in Arg,Glu, Lys, Arg/Glu, Arg/Lys, and Lys/Glu solutions. Increasing excipient concentration is denoted by increased shading (light to dark; left to right). Mean values are reported from three replicate REUS simulations. Error bars were estimated via error propagation (see SI for details).}
    \label{fig:pmf}
\end{figure*}

The free energy of hydrophobic polymer folding in excipient solutions is reported in (Fig. ~\ref{fig:pmf}). At 0.25 M concentrations, hydrophobic polymer collapse is favored in Arg solutions, while unfolding is favored in Lys and Glu solutions. At 0.5 M and 1.0 M concentrations, Arg, Lys, and Glu solutions individually favor hydrophobic polymer collapse, relative to pure water.  At this concentration, Arg is the most effective single excipient in stabilizing the folded hydrophobic polymer state. Lys is the next most effective excipient, while Glu is the least effective. Given the rank-ordering of individual excipients provided here (Arg\textgreater Lys\textgreater  Glu), positive amino acid charge (Arg, Lys) may be an important feature over negatively charged amino acids (Glu). Previous studies have similarly highlighted the integral role of ions in the salting-out phenomenon associated with hydrophobic interaction stabilization.\cite{zangi_effect_2007}

Among binary excipient mixtures, both Arg/Glu and Arg/Lys  significantly stabilized the folded polymer state, particularly at 1.0 M (Fig. ~\ref{fig:pmf}). Further, we consider 1.0 M Arg/Lys and Arg/Glu mixtures to be synergistic relative to their individual components, as the free energy of polymer folding in these solutions is more favorable than in any single excipient solution. In Lys/Glu mixtures, hydrophobic polymer folding is favored at 0.25 M, whereas the unfolded polymer is favored in Lys or Glu solutions individually at this concentration. Hence, we identify Lys/Glu mixtures to be synergistic as well at 0.25 M, although not at higher concentrations.

\subsection{Thermodynamic Components of Hydrophobic Polymer Collapse in Excipient Solutions}
\begin{figure*}[!ht]
   \includegraphics[width=1.0\textwidth]{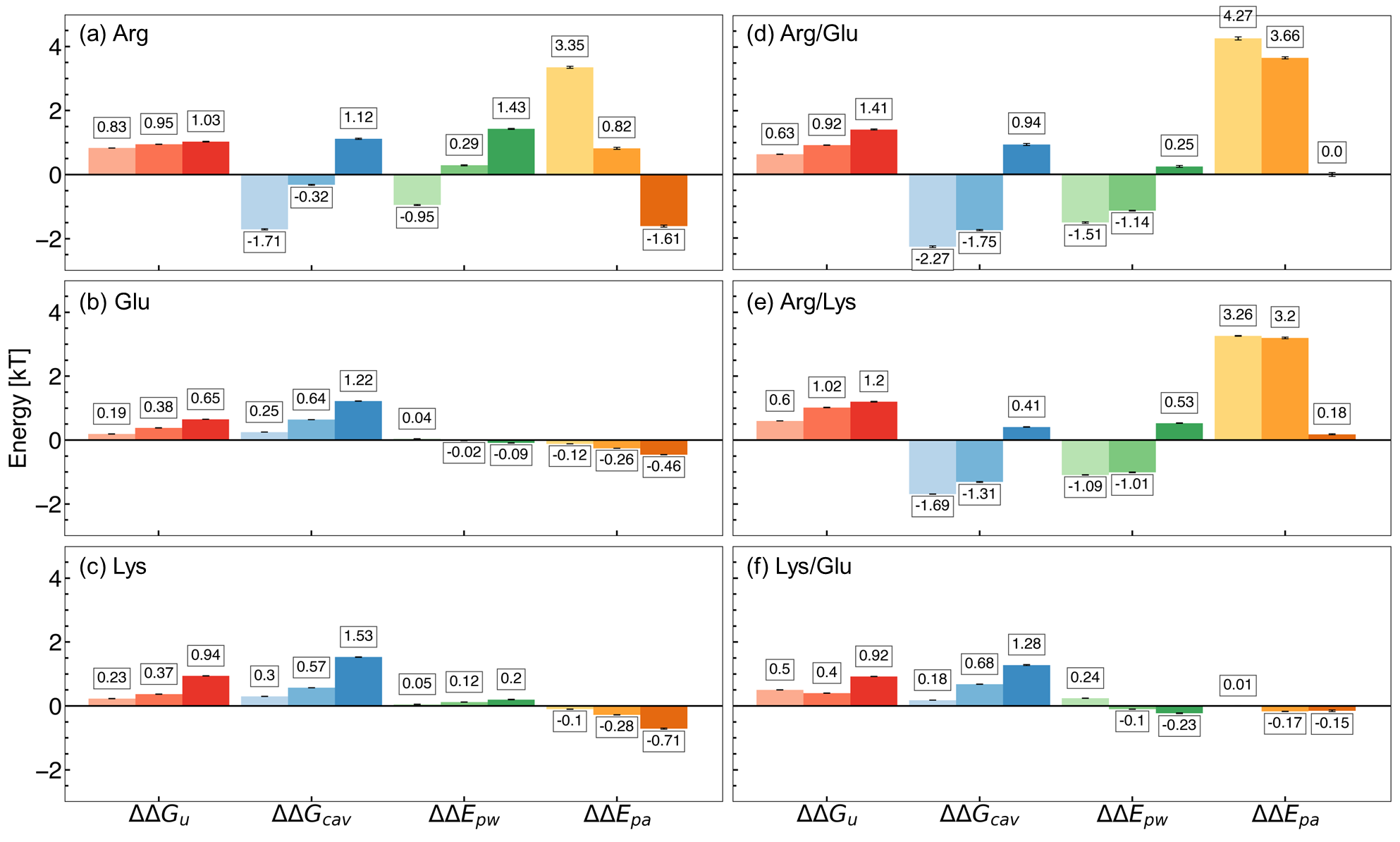}
    \caption{Contributions to the free energy of hydrophobic polymer unfolding in 0.25 M, 0.50 M, and 1.0 M (a) Arg, (b) Glu, (c) Lys, (d) Arg/Glu, (e) Arg/Lys, and (f) Lys/Glu solutions. Changes in overall free energy of unfolding ($\Delta \Delta G_{u}$), cavitation contribution ($\Delta \Delta G_{cav}$), polymer-water interactions ($\Delta \Delta E_{pw}$), and polymer-additive interactions ($\Delta \Delta E_{pa}$) are shown. Increasing additive concentration is denoted by increased shading (light to dark; left to right). Mean values are reported from three replicate REUS simulations. Error bars were estimated via error propagation (see SI for details).}
    \label{fig:dbars}
\end{figure*}

To explore the thermodynamic origins of excipient effects on hydrophobic polymer collapse, we decomposed the PMF into individual components. Fig. ~\ref{fig:dbars} shows the change in each component upon unfolding in excipient solution relative to that observed in water. The first $\Delta$ arises from the difference between folded and unfolded states (e.g., $\Delta E = \langle E_{u} \rangle - \langle E_{f} \rangle$), and the second $\Delta$ arises from the free energy difference between excipient solution ($\Delta E_{exc}$) and water ($\Delta E_{wat}$) (e.g., $\Delta\Delta E = \Delta E_{exc} - \Delta E_{wat}$). In all cases, we found $\Delta \Delta G_{vac} \sim 0$, as $W_{vac}$ does not depend on the solvent. Additionally, the change in polymer-counterion interaction energy, $\Delta \Delta E_{c}$,  was observed to be near 0 in all cases. Hence, these terms were omitted from Fig.~\ref{fig:dbars}, for clarity.

%% Results for Arg, PMF decomposition
As we have previously reported,\cite{Zajac2024arXiv} in Arg solutions, the favorability of individual components is concentration dependent (Fig.~\ref{fig:dbars}a). In particular, we observe that at low concentrations, direct polymer-Arg interactions favor collapse, while at high concentrations, the cavity component and polymer-water interactions drive folded state stability.

%% Results for glutamate and lysine, PMF decomposition
For  Glu (Fig.~\ref{fig:dbars}b) and Lys (Fig.~\ref{fig:dbars}c) solutions, a monotonic increase in the overall favorability of polymer collapse is observed with increasing concentration, which appears to be driven primarily by a favorable cavity component. At the same time, polymer-Glu and polymer-Lys interactions oppose collapse, while the polymer-water component is negligible.

%% Results for Arg-containing binary mixtures, PMF decomposition
For binary mixtures Arg/Glu (Fig.~\ref{fig:dbars}d) and Arg/Lys (Fig.~\ref{fig:dbars}e), direct polymer-additive interactions dominate the free energy of polymer collapse at 0.25 M and 0.5 M, while at 1.0 M, this component is negligible. In contrast, the cavity component and polymer-water interactions favor polymer unfolding at 0.25 M and 0.5 M, while these components favor polymer collapse at 1.0 M.

%% Results for lysine/glutamate mixture, PMF decomposition
In the case of Lys/Glu (Fig.~\ref{fig:dbars}f) solutions, stabilization of the folded polymer is nearly completely determined by the cavity component. A monotonic increase in this component is observed with increasing Lys/Glu concentration, while polymer-water and polymer-additive interactions are negligible.

\subsection{Excipient Synergy is Driven by Changes in Direct Interactions}
\begin{figure*}[!ht]
   \includegraphics[width=0.6\textwidth]{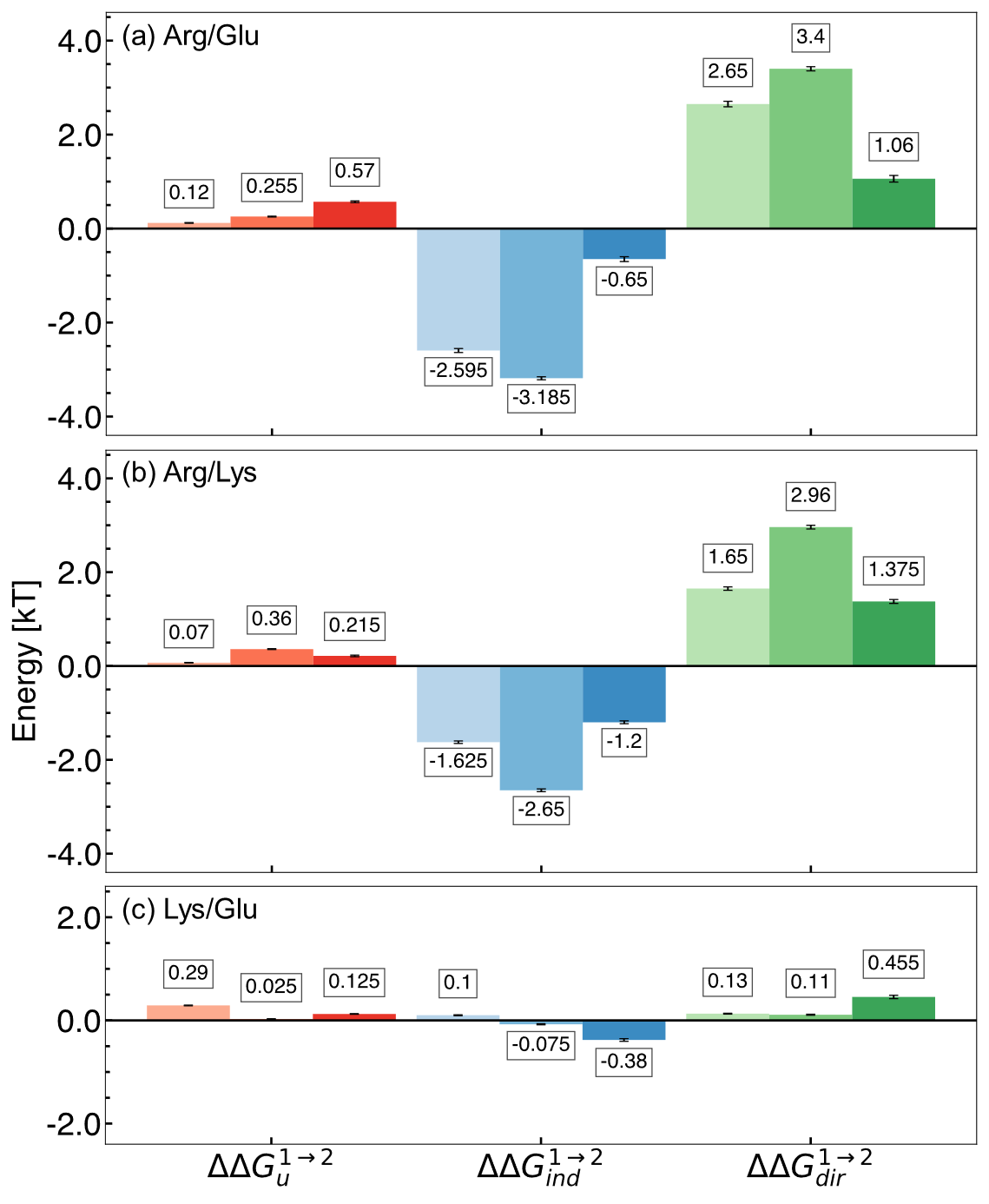}
    \caption{Excipient synergy observed in 0.25 M, 0.50 M, and 1.0 M (a) Arg/Glu, (b) Arg/Lys, and (c) Lys/Glu solutions. Changes in overall free energy of unfolding ($\Delta \Delta G_{u}^{1 \rightarrow 2}$), direct interactions ($\Delta \Delta G_{dir}^{1 \rightarrow 2}$), and indirect interactions ($\Delta \Delta G_{ind}^{1 \rightarrow 2}$) and polymer are shown. $1 \rightarrow 2$ denotes the difference associated with changing from two single excipient solutions to a binary excipient solution. Increasing additive concentration is denoted by increased shading (light to dark; left to right). Mean values are reported from three replicate REUS simulations. Error bars were estimated via error propagation (see SI for details).}
    \label{fig:dbars-sum}
\end{figure*}

%% Exploring excipient synergy, ddG 1->2 metric
To quantify the extent of synergy in binary excipient mixtures, we computed the unfolding free energy difference associated with changing from the average of two single-excipient solutions to one binary-excipient solution ($\Delta \Delta G_{u}^{1 \rightarrow 2}$), at a given total excipient concentration. This quantity is computed as:
\begin{equation}\label{dG_synergy}
     \Delta \Delta G_{u}^{1 \rightarrow 2} = \Delta G_{u}^{a,b} - \left(\frac{\Delta G_{u}^{a}+\Delta G_{u}^{b}}{2}\right)
\end{equation}
for excipients $a$ and $b$.  In cases where $\Delta \Delta G_{u}^{1 \rightarrow 2} > 0$, the binary excipient mixtures have a more favorable effect on hydrophobic polymer collapse than did the individual components. A mechanistic understanding of this synergy may be obtained by detailing the balance of interactions involved. Specifically, we investigate both direct interactions (polymer-additive and polymer-counterion; $\Delta\Delta G_{dir}$) and indirect effects (cavitation and polymer-water; $\Delta\Delta G_{ind}$) associated with each solution (Fig.~\ref{fig:dbars-sum}).

%Describe results of ddG_u^1->2
In all binary excipient solutions, we observe favorable changes in the free energy of polymer folding, relative to single excipient solutions. In Arg/Glu (Fig.~\ref{fig:dbars-sum}a) and Arg/Lys (Fig.~\ref{fig:dbars-sum}b) solutions, this increased favorability is associated with a favorable change in direct polymer-additive interactions. At the same time, folded state stability is opposed by unfavorable indirect components, relative to in single excipient solutions. The same observations are made for Lys/Glu (Fig.~\ref{fig:dbars-sum}c) solutions, albeit to a lesser extent. Interestingly, a different optimal concentration (with maximal $\Delta \Delta G_{u}^{1 \rightarrow 2}$) is observed for each pair of excipients -- 1.0 M for Arg/Glu, 0.5 M for Arg/Lys, and 0.25 M for Lys/Glu.

%Brief discussion
The manifestation of the observed synergy describes a mechanism for improving the effectiveness of Arg-containing solutions. We observe that, while Arg is effective in stabilizing hydrophobic polymer collapse, attractive polymer-Arg interactions drive unfolding at 1.0 M concentration. In the presence of Lys or Glu, this opposition to collapse is eliminated. Hence, co-excipient addition is a suitable strategy to manipulate the underlying mechanism of Arg in hydrophobic polymer collapse.

\begin{figure*}[!ht]
   \includegraphics[width=1.0\textwidth]{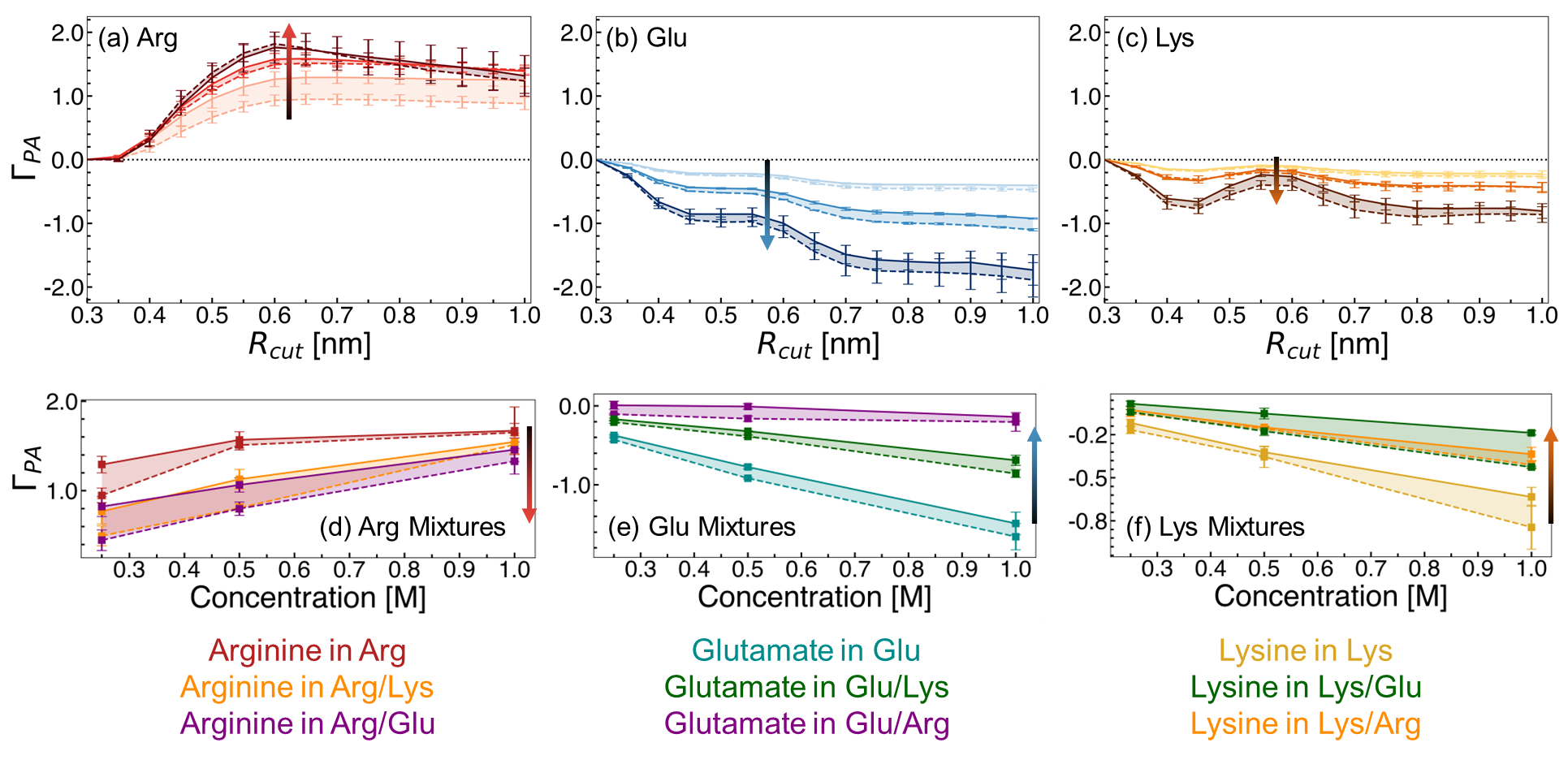}
    \caption{Preferential interaction coefficient values as a function of the cut-off distance for the local domain of the hydrophobic polymer for (a) Arg, (b) Glu, and (c) Lys. Changes in $\Gamma_{PA}$ at $R_{cut} = 0.7$ nm are shown for (d) Arg mixtures, (e) Glu mixtures, and (f) Lys mixtures. Dashed lines indicate values for the unfolded state, while solid lines denote the folded state. Increasing concentration is denoted by increased shading (light to dark). Arrows denote the trend with increasing concentration in a-c, while marking changes following co-excipient addition in d-f. Mean values and errors were estimated from three replicate simulations. Errors are reported as standard deviations from mean values.}
    \label{fig:prefInt}
\end{figure*}

We hypothesize that the primary source for co-excipient synergy results in a change in the balance of polymer-water-excipient interactions. Preferential interaction coefficients, $\Gamma_{PA}$, is a powerful metric for quantifying this balance. $\Gamma_{PA} > 0$ indicates relative accumulation of an excipient in the local polymer domain, while $\Gamma_{PA} < 0$ indicates relative depletion. In Fig.~\ref{fig:prefInt}, we report $\Gamma_{PA}$ for Arg, Lys, and Glu solutions. This analysis reveals that Arg preferentially interacts with the hydrophobic polymer (Fig.~\ref{fig:prefInt}a), while Glu (Fig.~\ref{fig:prefInt}b) and Lys (Fig.~\ref{fig:prefInt}c) are preferentially excluded. Further, these trends were observed to increase with concentration. 

%Preferential interaction coefficient, binary excipients
We explored the change in excipient distribution by considering $\Gamma_{PA}$ of an excipient when alone versus in a binary excipient solution. In Fig.~\ref{fig:prefInt}d-f, we show ensemble averaged $\Gamma_{PA}$ values using an $R_{cut}$ value of 0.7 nm. This value is selected as a cut-off distance because, beyond this distance, no significant changes in $\Gamma_{PA}$ are observed for the excipients. From this perspective, the preferential accumulation of Arg near the polymer is reduced in Arg/Lys or Arg/Glu solutions, relative to in Arg solutions alone (Fig. \ref{fig:prefInt}d). We have previously highlighted that, at 1.0 M Arg concentration, direct polymer-Arg interactions drive unfolding.\cite{Zajac2024arXiv}  Hence, we hypothesize that reduction in polymer-Arg interactions upon Lys or Glu incorporation in the solution results in net stabilization of the folded hydrophobic polymer.

Alongside the changes in Arg distribution, the relative accumulation of Lys or Glu is increased in Arg/Lys and Arg/Glu solutions, relative to in single excipient solutions (Fig. \ref{fig:prefInt}e,f). Overall, these findings imply a mutual recruitment of co-excipient A into the preferred domain of co-excipient B.

%Comparison of pref int to ddG_u^(1->2)
The change in $\Gamma_{PA}$ observed in binary excipient solutions describes, in part, the effects observed in Fig. ~\ref{fig:dbars-sum}. In Arg/Glu and Arg/Lys solutions, there is a depletion in $\Gamma_{PA}^{Arg}$ relative to in Arg solutions alone, resulting in a net reduction of polymer-additive interactions. Correspondingly, a favorable change in the direct component, $\Delta \Delta G_{u}^{1 \rightarrow 2}$, arises in Arg/Glu and Arg/Lys solutions, conferring increased stabilization of hydrophobic polymer collapse.

\subsection{Stabilizing Co-Excipients Preserve the Network Effects of Arg}
\begin{figure*}[!ht]
   \includegraphics[width=1.0\textwidth]{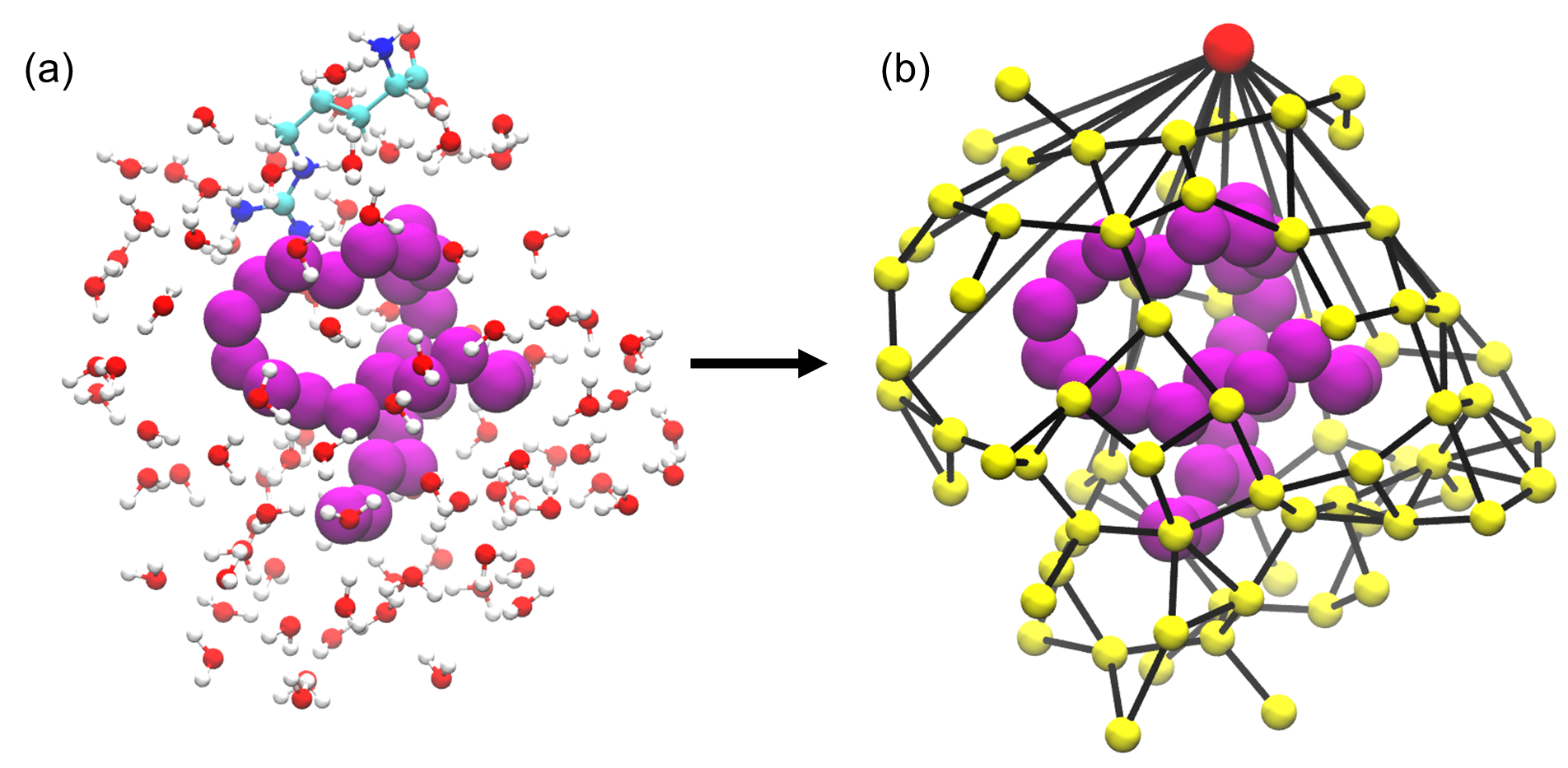}
    \caption{Graph/network representation of the hydrophobic polymer local environment. (a) Representative snapshot taken from an REUS trajectory. (b) Graph representation of the snapshot in (a). The hydrophobic polymer, which is not included in the graph but is added here for illustration, is represented by purple spheres. Water nodes in the local polymer domain are colored yellow, while an excipient node is colored in red. Edges between connecting nodes are drawn as black lines.}
    \label{fig:polymer-network}
\end{figure*}

Networks of excipient-water interactions embedding the hydrophobic polymer were analyzed using principles of network theory. In this approach, we treat the center-of-mass of all excipient and water molecules within 0.7 nm of the polymer as nodes, while edges are constructed between nodes where any pair of heavy atoms are located within 0.4 nm of one another (Fig. ~\ref{fig:polymer-network}). To quantify the flow of molecular interactions within the solvent network, we measured closeness centrality, $C_{c}$, and betweenness centrality, $C_{b}$. \textit{Closeness} centrality can be regarded as a measure of how long it takes to spread information from node $u$ to all other nodes sequentially. Correspondingly, this quantity is a measure of how close a node is to the center of the network. \textit{Betweenness} centrality, on the other hand, quantifies the number of times a node acts as a bridge along the shortest path between two other nodes. In other words, this quantity measures the propensity for a node to act as a "hub" of information propagation. For both measurements, we resolve centrality from all water or excipient nodes.

\begin{figure*}[!ht]
   \includegraphics[width=1.0\textwidth]{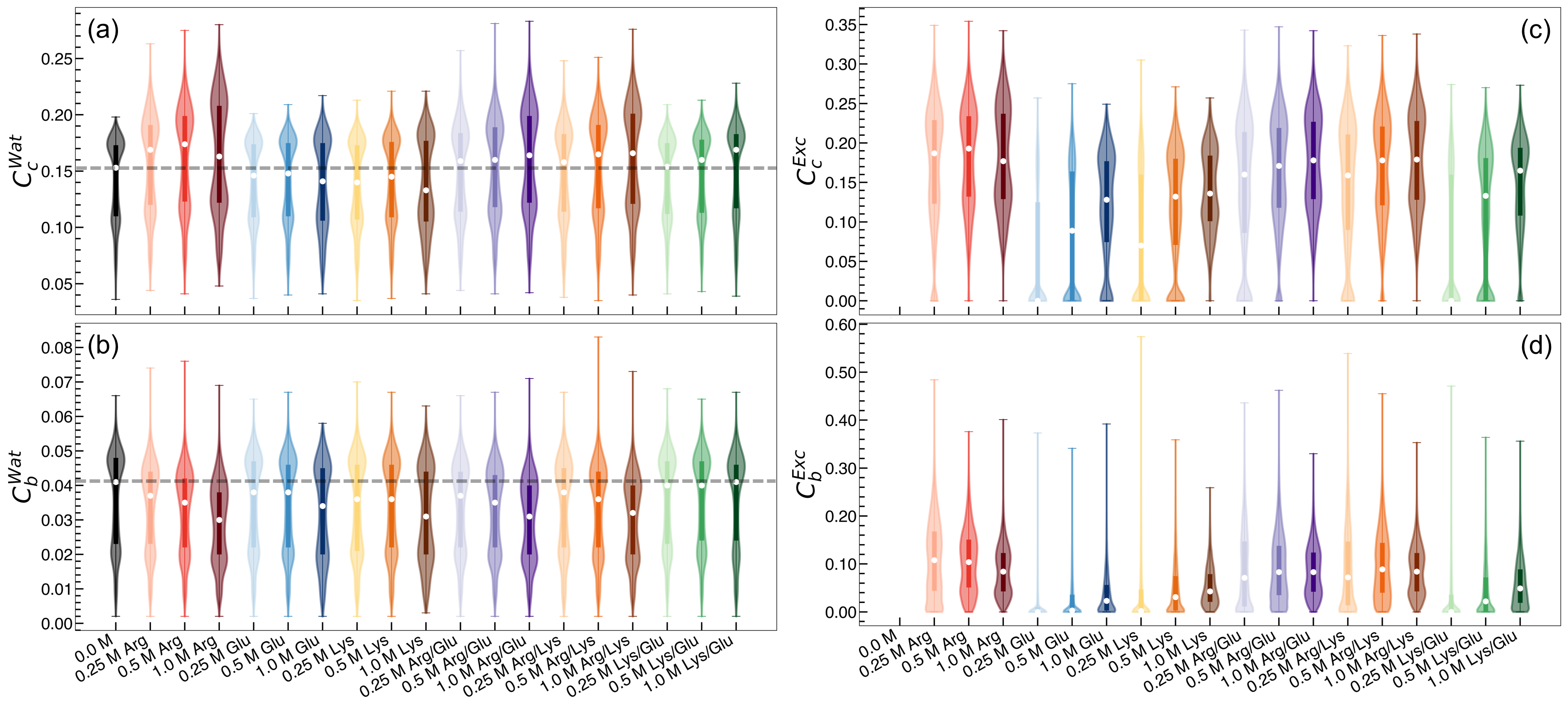}
    \caption{Violin plots of centrality measurements for the network representing the local polymer environment. (a) Closeness centrality among water nodes. (b) Betweenness centrality among water nodes. (c) Closeness centrality among excipient nodes. (d) Betweenness centrality among excipient nodes. Mean values are denoted by white dots. In a and b, the horizontal dashed line is used to mark mean centrality values for pure water.}
    \label{fig:centrality}
\end{figure*}

Distributions of centrality measurements obtained from folded state configurations are shown in Fig.~\ref{fig:centrality}. Distributions from unfolded configurations result in qualitatively similar trends (Fig. S4, Fig. S5), and hence were omitted from Fig.~\ref{fig:centrality}, for clarity. In Arg solutions, we observe an increase in water closeness centrality ($C_{c}^{Wat}$) (Fig.~\ref{fig:centrality}a) and a decrease in water betweenness centrality ($C_{b}^{Wat}$)  (Fig.~\ref{fig:centrality}b). The increase in $C_{c}^{Wat}$ indicates shorter distances from water nodes to all other nodes -- in other words, in Arg solutions, connectivity is increased among water molecules. This observation supports previous attempts to describe excipient effects via a network-based approach, which found that proteins in solution with stabilizing excipients have more compact interaction networks, relative to those in the presence of denaturants.\cite{miotto_osmolyte-induced_2023}

On the other hand, the decrease in $C_{b}^{Wat}$  suggests less information is transferred through water molecules. Correspondingly, relatively high excipient betweenness ($C_{b}^{Exc}$) distributions are observed for Arg solutions (Fig. ~\ref{fig:centrality}d). Such a finding conveys that Arg molecules integrate well into the local polymer environment, acting as central hubs for information transfer in the local solvent structure. Elsewhere, it has been reported that stabilizing osmolytes have similarly high betweenness centrality values,\cite{sundar_unraveling_2021} implying, together with our results, that integration into the local solvent environment with minimal disruption to the solvent interaction network may be a general phenomenon for excipient stability.

Relative to the centrality measurements associated with Arg solutions, network analysis reveals several key differences between solutions containing Lys or Glu. In solutions containing Lys or Glu as the only excipients, both $C_{c}^{Wat}$ and $C_{b}^{Wat}$ are found to decrease, or at least resulted in negligible changes relative to pure water. This finding suggests that water molecules become less connected within the local environment network (Fig. ~\ref{fig:centrality}a), while also becoming less essential hubs for information propagation (Fig. ~\ref{fig:centrality}b). In contrast to Arg solutions, $C_{c}^{Exc}$ and $C_{b}^{Exc}$ are both relatively low, implying that Lys and Glu do not integrate into the local solvent environment as well as Arg does (Fig. ~\ref{fig:centrality}c,d).

These centrality measurements reflect strongly connected networks in Arg solutions, and networks with poor connectivity in Lys or Glu solutions. In binary excipient solutions containing Arg (Arg/Lys and Arg/Glu), we observe an overall preservation of the strong connectivity identified in Arg solutions alone. Specifically, $C_{c}^{Wat}$ is found to increase in Arg/Lys and Arg/Glu solutions relative to pure water, reflecting increased connectivity among water molecules (Fig. ~\ref{fig:centrality}a). Similar to Arg solutions, $C_{b}^{Wat}$ is observed to decrease with a concomitant increase in $C_{b}^{Exc}$, reflecting favorable integration of excipient molecules into the local polymer environment in Arg/Lys and Arg/Glu solutions (Fig. ~\ref{fig:centrality}b,d). Finally, $C_{c}^{Exc}$ is consistently higher in Arg/Lys and Arg/Glu solutions relative to Lys or Glu alone, marking an increase in local excipient connectivity in these binary excipient solutions (Fig. ~\ref{fig:centrality}c). In general, the Lys/Glu binary excipient solution reflected network properties similar to Lys or Glu solutions alone.

\begin{figure*}[!ht]
   \includegraphics[width=1.0\textwidth]{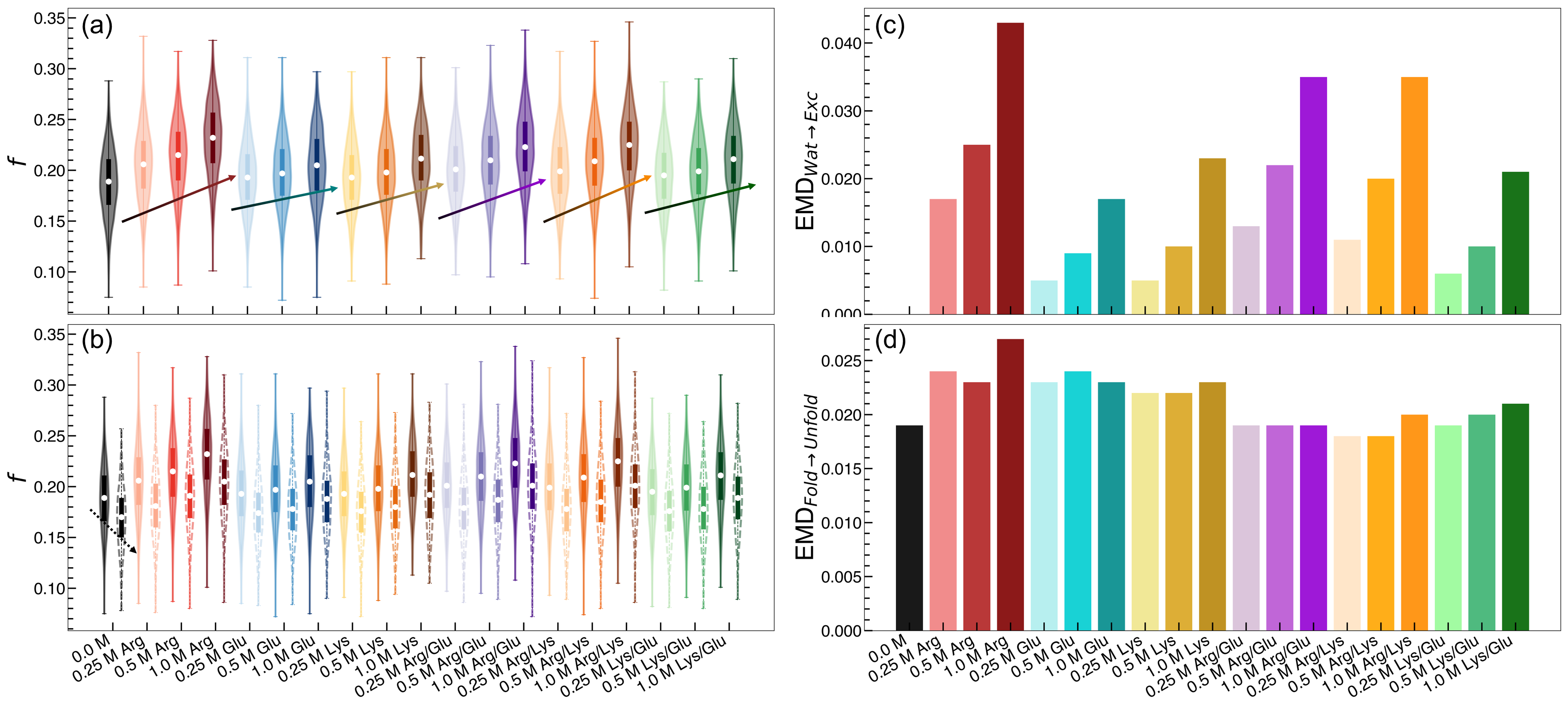}
    \caption{Graph fragmentation analysis of the local hydrophobic polymer environment. (a) Fragmentation threshold, $f$, for all solutions. Arrows are drawn to guide changes with increasing concentration for a given excipient solution. (b) Fragmentation threshold distributions split between folded (solid) and unfolded (dashed) states. An arrow is drawn for the 0.0 M solution to guide changes in distribution upon polymer unfolding. (c) Earth mover's distance (EMD) between the 0.0 M fragmentation threshold distribution in a and a given excipient distribution. (d) EMD between folded and unfolded state fragmentation threshold distributions for all solutions.}
    \label{fig:fragmentation}
\end{figure*}

We hypothesize that solutions with stronger connectivity in the local polymer environment may confer greater stability to the network itself. To assess this, we compute a fragmentation metric, $f$, which measures the average number of nodes that must be removed to form two independent, disconnected graphs. Such an approach is inspired by percolation theory, and has been used elsewhere to measure the stability of large biomolecular assemblies, including viral capsids.\cite{brunk_molecular_2018, brunk_percolation_2021} To better quantify differences in distribution, we computed the Earth Mover's Distance (EMD)\cite{rubner_earth_2001} as a measure of distribution dissimilarity, where higher EMD values indicate less overlap between two distributions. On average, local polymer environments are more resistant to fragmentation in Arg, Arg/Lys, and Arg/Glu solutions (Fig.~\ref{fig:fragmentation}a,c), as indicated by increasing $f$ distributions and relatively large EMD values. For all solutions, $f$ values are higher in the local environments of folded polymer states, relative to unfolded polymer states (Fig.~\ref{fig:fragmentation}b). However, this appears to be a general trend, and the differences between folded state and unfolded state graph stability does not depend on solution identity (Fig.~\ref{fig:fragmentation}d).

\subsection{Rationalizing Co-Excipient Synergy}
Overall, we have uncovered that Arg stabilizes hydrophobic polymer collapse to a greater extent than Lys, Glu, or Lys/Glu solutions. Further, addition of Lys or Glu to Arg solutions give rise to synergistic effects, as the resulting Arg/Lys and Arg/Glu solutions are the most effective stabilizing solutions we have investigated. Further, these synergies are associated with stabilizing indirect effects at high concentration and a dramatic reduction in destabilizing direct interactions. Correspondingly, the solvent distribution, as characterized by preferential interaction coefficient analysis, reflects a reduction in Arg accumulation in the local hydrophobic polymer domain.

Our network analysis implies that Arg integrates well into the local polymer environment, increases connectivity among water molecules, and increases the stability of the solvent network embedding the hydrophobic polymer. Importantly, these key network properties persist for Arg solutions to a greater extent than Lys or Glu solutions. Moreover, addition of Lys or Glu to Arg solution (Arg/Lys, Arg/Glu) preserves the network properties of Arg alone, which may be a key aspect giving rise to the observed synergy among these excipients. 

We hypothesize that addition of Lys or Glu to Arg solutions results in the formation of the same stable, highly connected local polymer environment as in solutions containing only Arg, while simultaneously reducing the penalty associated with direct polymer-Arg interactions. While our findings support the hypothesis described above, we aimed to uncover key molecular features or signatures that drive these changes in polymer stability. To this end, we carried out additional analyses to explore whether Lys and Glu undergo unique molecular mechanisms to arrive at similar synergistic effects with Arg.

\subsubsection{Rationalizing Arg/Lys Synergy: The Sticky Guanidinium Hypothesis}
Given the importance of excipient-excipient interactions in dictating solution structure, we aimed to explore these molecular interactions further. In particular, clustering has been implicated as a key feature associated with the excipient effects of Arg.\cite{das_inhibition_2007, schneider_investigation_2009, shukla_interaction_2010, schneider_arginine_2011, vagenende_protein-associated_2013, santra_analyzing_2021, santra_influence_2022, meingast_physiochemical_2021} Among amino acid excipients, these molecules interact in solution via three primary modes of interaction: (i) backbone-backbone (COO$^{-}$-NH$_{3}^{+}$), (ii) backbone-sidechain (Gdm$^{+}$-COO$^{-}$), and (iii) sidechain-sidechain (Gdm$^{+}$-Gdm$^{+}$). Using geometric criteria for identifying these specific interactions, we carried out an analysis of excipient cluster formation in different excipient solutions.

\begin{figure*}[!ht]
   \includegraphics[width=1.0\textwidth]{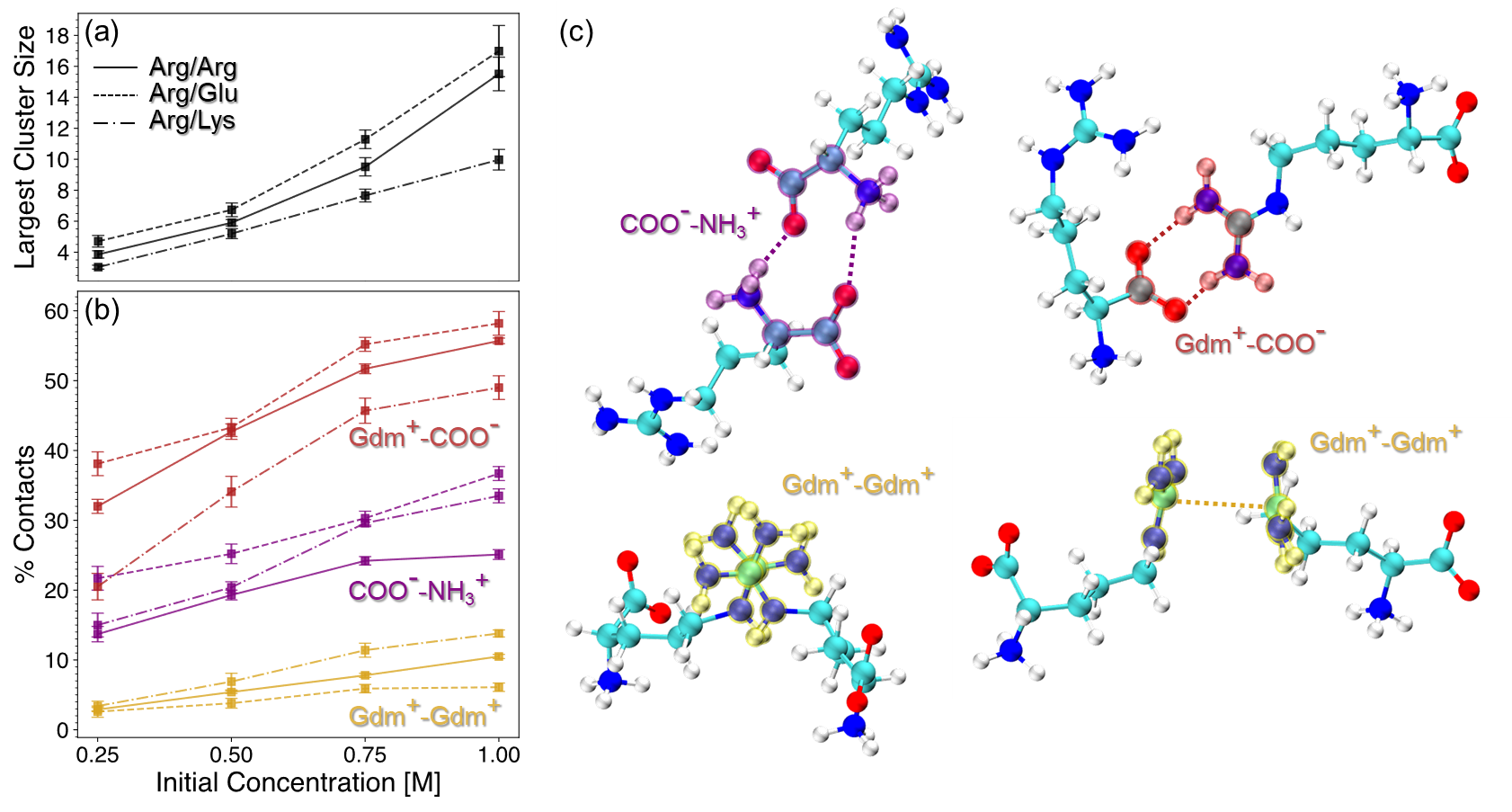}
    \caption{Excipient clustering analysis. (a) Largest cluster size observed from excipient solutions as a function of concentration. (b) Average percentage of each specific contact observed in excipient clusters for different solutions. (c) Representative configurations showing the three primary modes of excipient-excipient interactions. In (a) and (b), solid lines represent solutions containing only Arg, dashed represents Arg/Glu, and dotted dashed lines represent Arg/Lys. In (b) and (c), different colors represent COO$^{-}$-NH$_{3}^{+}$ (purple), Gdm$^{+}$-COO$^{-}$ (red), and Gdm$^{+}$-Gdm$^{+}$ (yellow) interaction types.}
    \label{fig:clustering}
\end{figure*}

Overall, the extent of cluster formation, as measured by the average largest cluster size, is greater in Arg and Arg/Glu solutions than in Arg/Lys (Fig. ~\ref{fig:clustering}a). We attribute this finding to favorable electrostatic interactions between Arg and Glu, as well as the unique properties of the Gdm$^{+}$ sidechain of Arg that enables favorable like-charge interactions.\cite{gund_guanidine_1972, mason_hydration_2003, shukla_interaction_2010, shih_cation-cation_2013, vazdar_arginine_2018} The presence of cluster formation in all Arg-containing solutions is primarily driven by Gdm$^{+}$-COO$^{-}$ interactions (Fig. ~\ref{fig:clustering}b,c), a finding consistent with previous work detailing the importance of Arg "head-to-tail" stacking.\cite{shukla_interaction_2010} Interestingly, while excipient clustering is reduced overall in Arg/Lys solutions relative to Arg alone, the extent of Gdm$^{+}$-Gdm$^{+}$ pairing is increased in Arg/Lys solutions.

\begin{figure*}[!ht]
   \includegraphics[width=1.0\textwidth]{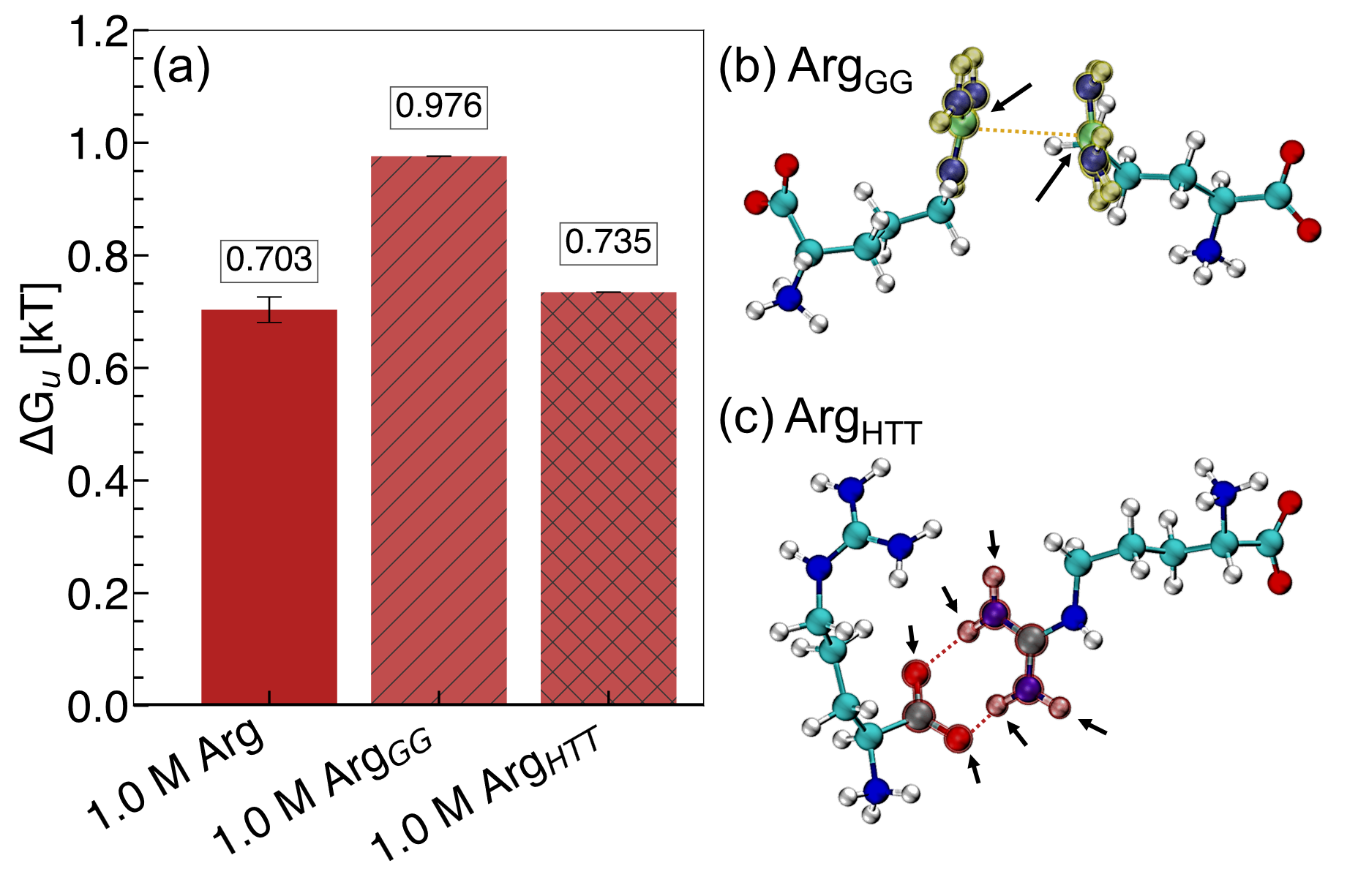}
    \caption{Free energy of hydrophobic polymer collapse in modified 1.0 M Arg solutions. (a) $\Delta G_{u}$ for unmodified Arg, increased increased Gdm$^{+}$-Gdm$^{+}$ interaction strength (Arg$_{GG}$), and increased Gdm$^{+}$-COO$^{-}$ interaction strength (Arg$_{HTT}$). (b) Modified interactions in Arg$_{GG}$. (c) Modified interactions in Arg$_{HTT}$. In (b) and (c), arrows denote atoms with scaled interaction parameters.}
    \label{fig:gdmhtt}
\end{figure*}

To probe the importance of these interactions relevant to excipient clustering, we carried out two additional REUS simulations of hydrophobic polymer folding with modified Arg-Arg interaction parameters (Fig.~\ref{fig:gdmhtt}). To simulate increased Gdm$^{+}$-Gdm$^{+}$ pairing among Arg molecules (Arg$_{GG}$), we scaled the interaction strength between Gdm$^{+}$ carbons by 150\% (Fig.~\ref{fig:gdmhtt}b). Similarly, to simulate increased Gdm$^{+}$-COO$^{-}$ head-to-tail pairing among Arg molecules (Arg$_{HTT}$), the interaction strength between COO$^{-}$ oxygens and Gdm$^{+}$ hydrogens was scaled by 150\% (Fig.~\ref{fig:gdmhtt}c). From these simulations, the free energy of polymer folding in Arg$_{GG}$ solution becomes significantly more favorable relative to unmodified Arg solution, while Arg$_{HTT}$ results in only a small increase in folded state stability (Fig.~\ref{fig:gdmhtt}a).

These results demonstrate the importance of the Gdm$^{+}$ sidechain of Arg in hydrophobic polymer collapse. Gdm$^{+}$ is a known protein denaturant, and elsewhere, has been shown to drive unfolding of an elongated hydrophobic polymer at high concentrations.\cite{godawat_unfolding_2010} Further, we have shown previously that direct interactions between the Gdm$^{+}$ sidechain of Arg and the hydrophobic polymer favors polymer folding at lower concentrations.\cite{Zajac2024arXiv} Overall, our mechanistic explanation for Arg/Lys synergy involves a Lys-mediated increase in Gdm$^{+}$-Gdm$^{+}$ "stickiness" among Arg molecules. We hypothesize that upon addition of Lys, the increase in Gdm$^{+}$-Gdm$^{+}$ pairing among Arg molecules gives rise to the favorable change in $\Delta\Delta E_{pa}$ observed in Arg/Lys solutions. This is achieved by limiting the number of Gdm$^{+}$ interaction sites available to interact with the polymer, resulting in a relative depletion of Arg from the local polymer domain. 

\subsubsection{Rationalizing Arg/Glu Synergy: The Dynamics Reducing Hypothesis}
While Arg/Lys synergy appears to be associated with changes in Gdm$^{+}$-Gdm$^{+}$ pairing among Arg molecules, we did not observe the same changes in excipient clustering in Arg/Glu solutions. Hence, to explain molecular-level changes linked to Arg/Glu synergy, we turned our attention to the behavior of water molecules in the local polymer environment. 

\begin{figure*}[!ht]
   \includegraphics[width=1.0\textwidth]{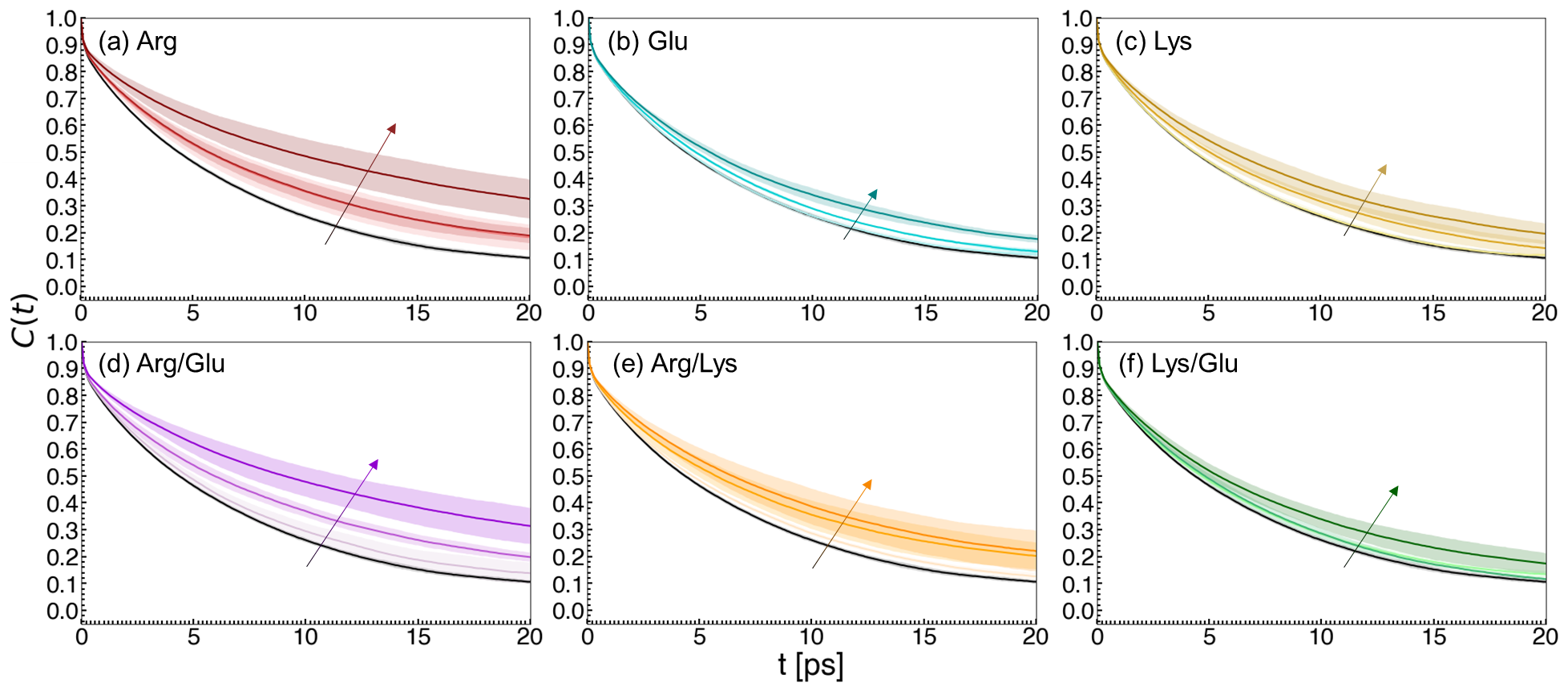}
    \caption{Water reorientation dynamics in the local hydrophobic polymer domain. The time correlation function, $C(t)$, of instantaneous water dipole moments are plotted for (a) Arg, (b) Glu, (c) Lys, (d) Arg/Glu, (e) Arg/Lys, and (f) Lys/Glu solutions. Increasing excipient concentration is shown by increased shading, while values obtained in pure water are represented by black curves. Arrows are drawn to guide the change with concentration.}
    \label{fig:watdyn}
\end{figure*}

To this end, we computed water reorientation dynamics in the local hydrophobic polymer domain for our excipient solutions (Fig. ~\ref{fig:watdyn}). The characteristic reorientation time for each solution was computed by fitting water dipole correlation functions to an exponential decay. With increasing excipient concentration, the reorientation time ($\tau$) of local water molecules was observed to increase, indicating a slowing of water dynamics (Table ~\ref{tbl:tau}). This effect is most pronounced in Arg and Arg/Glu solutions.

\begin{table*}
\small
  \caption{Reorientation times (ps) of the instantaneous dipole vector of water molecules local to the hydrophobic polymer.}
  \label{tbl:tau}
  \begin{tabular}{c c c c c c c}
    \hline
    Concentration (M) & Arg & Lys & Glu & Arg/Glu & Arg/Lys & Lys/Glu \\
    \hline
    0.0 M & 6.54 (0.02) & - & - & - & - & - \\
    0.25 M & 8.52 (1.2) & 6.46 (0.1) & 7.07 (0.5) & 7.24 (0.1) & 6.9 (0.1) \\
    0.5 M & 8.64 (0.4) & 7.85 (0.5) & 7.17 (0.1) & 8.86 (0.4) & 8.56 (1.4) & 7.17 (0.3) \\
    1.0 M & \textbf{12.6 (1.0)} & 9.29 (1.4) & 8.4 (0.4) & \textbf{12.37 (0.8)} & 8.81 (0.9) & 8.48 (0.6) \\
    \hline
  \end{tabular}
\end{table*}

Similar reductions in hydration shell dynamics have been associated with an increase in melting temperature of proteins. A recent study has proposed that stabilizing osmolytes slow down the dynamics of water, while denaturants accelerate water dynamics, inducing a pseudo-temperature change experienced by the protein.\cite{hishida_effect_2022} Other studies have highlighted that osmolytes increase hydrogen bond relaxation time among water molecules and reduce rotational, translational, and tumbling motions of water.\cite{jas_reorientation_2016, diaz_effect_2023, saladino_simple_2011, jahan_conformational_2020, zeman_effect_2020, gazi_conformational_2023}

We hypothesize that the key consequence associated with this phenomenon is the formation of a rigid solvent network embedding the hydrophobic polymer in Arg and Arg/Glu solutions. In the case of Arg/Glu, Glu may provide an advantage relative to Arg alone due to a reduction of Gdm$^{+}$ sidechain accumulation in the local polymer domain, as reflected by $\Gamma_{PA}$. This results in the favorable change in $\Delta\Delta E_{pa}$, similar to Arg/Lys, while retaining the reduced water dynamics and stable local network associated with Arg solutions.

\section{Conclusions}
A growing body of literature has contributed to understanding the effects of Arg on biomolecular stability. In some cases, addition of co-excipients to formulations that include Arg has resulted in reversals\cite{anumalla_counteracting_2019} or synergistic\cite{shukla_understanding_2011} effects. Recently, we demonstrated that the peculiar placement of Arg at the edge of a mechanistic flip between indirect- and direct-dominated effects on hydrophobic interactions may be a key feature that enables tunable properties of Arg.\cite{Zajac2024arXiv} Our findings here show that, not only is Arg a more effective stabilizer of hydrophobic interactions than its Lys or Glu counterparts, but addition of these less effective excipients augments the effectiveness of Arg solution stability.

We observed that the primary mechanism associated with Arg/Glu and Arg/Lys synergy is a substantial reduction in direct polymer-excipient interactions that oppose collapse at high Arg concentration. Through preferential interaction coefficient analysis, we further identified that Lys and Glu are both effective at reducing the relative accumulation of Arg molecules in the local polymer domain, providing a mechanistic explanation for the favorable change in $\Delta\Delta E_{pa}$.

Analysis of the solvent network embedding the hydrophobic polymer provides an explanation for how excipients alter the local polymer environment. We found that Arg increases connectivity among water molecules, integrates favorably into the local environment, and increases the stability of the network by delaying the onset of simulated graph fragmentation. These features are more pronounced in Arg solutions than in Lys, Glu, or Lys/Glu solutions. Importantly, we identified these same features in Arg/Lys and Arg/Glu solutions, indicating that stabilizing co-excipients preserve the network effects of Arg.

Finally, we derived two hypotheses for Arg/Lys and Arg/Glu synergy. In the case of Arg/Lys, there is an increase in Gdm$^{+}$-Gdm$^{+}$ pairing among Arg molecules, potentially reducing the number of available Gdm$^{+}$ sidechains that drive polymer unfolding at high concentrations. From simulations of increased Gdm$^{+}$-Gdm$^{+}$ interaction strength between Arg molecules, we found this change to be sufficient for increasing the favorability of polymer collapse. In the case of Arg/Glu, we did not find increased Gdm$^{+}$-COO$^{-}$ paring in solution was sufficient to drive co-excipient synergy. However, we observed a reduction in the dynamics of water molecules local to the hydrophobic polymer in these solutions. Similar reductions in local water dynamics has been linked to increased melting temperature of proteins in the presence of stabilizing osmolytes.\cite{jas_reorientation_2016, diaz_effect_2023, saladino_simple_2011, jahan_conformational_2020, zeman_effect_2020, gazi_conformational_2023}

Here, we demonstrated that changes in excipient composition alter hydrophobic interactions, the dominant force associated with several biologically-important processes including protein folding and self-assembly. As it pertains to formulations, this is an important factor in preventing protein denaturation and aggregation. Due to its placement on the edge of a mechanistic flip, formulations containing Arg as an excipient may be improved by shifting the balance of direct- and indirect-mediated effects. To this end, we increased the relative stability of hydrophobic polymer collapse by reducing destabilizing direct effects via co-excipient addition of either Lys or Glu. Overall, these results highlight the investigation of molecular-level insights of excipient mechanisms as an important endeavor in the rational design of stable biologics.

\section*{Conflicts of interest}
There are no conflicts to declare.

\section*{Acknowledgements}
This material is based upon work supported by the National Science Foundation under DMREF Grant Nos. 2118788, 2118693, and 2118638. Computing resources were provided by the Minnesota Supercomputing Institute (MSI).

\bibliography{ref}

\end{document}

\setcounter{figure}{0}
\emergencystretch 3em

\begin{table*}
  \caption{Setup of simulated systems.}
  \label{tbl:si-simsetup}
  \begin{tabular}{c c c c}
    \hline
    System & Simulation Time (ns) & Concentration (M) & $N_{Exc}$ \\
    \hline
    Excipient & 20 & 0.25 & 47 \\
    Excipient & 20 & 0.50 & 93 \\
    Excipient & 20 & 1.0 & 185 \\
    Polymer & 100 x 12 & 0.00 & 0 \\
    Polymer + Excipient & 3 x 100 x 12 & 0.25 & 47 \\
    Polymer + Excipient & 3 x 100 x 12 & 0.50 & 93 \\
    Polymer + Excipient & 3 x 250 x 12 & 1.0 & 185 \\

    \hline
  \end{tabular}
\end{table*}

\begin{figure*}[!ht]
\centering
   \includegraphics[width=1.0\textwidth]{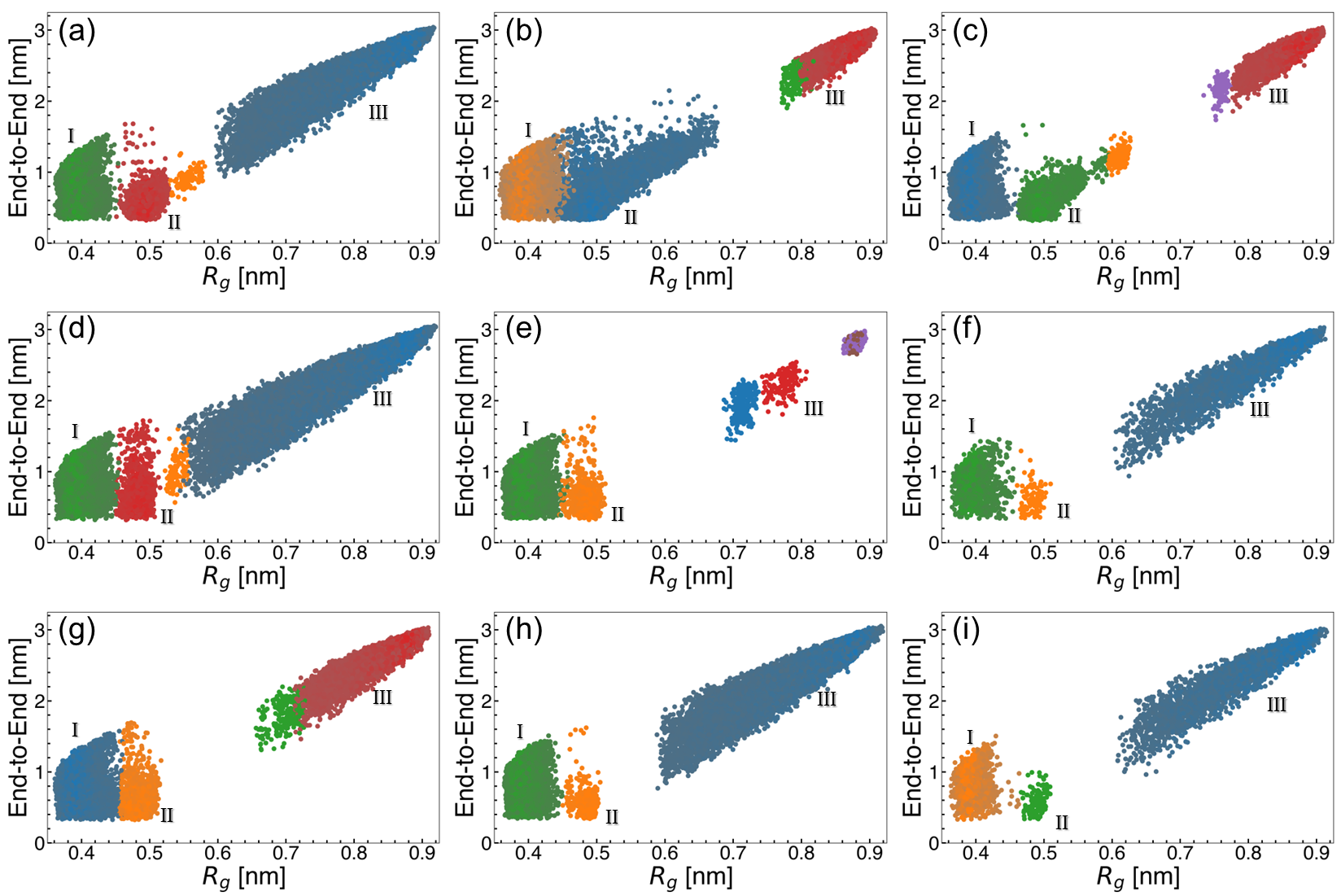}
   \caption{HDBSCAN clustering of polymer configurations for (a-c) Arg, (d-f) Glu, and (g-i) Lys solutions.} 
    \label{fig:si-HDBSCAN1}
\end{figure*}

\begin{figure*}[!ht]
\centering
   \includegraphics[width=1.0\textwidth]{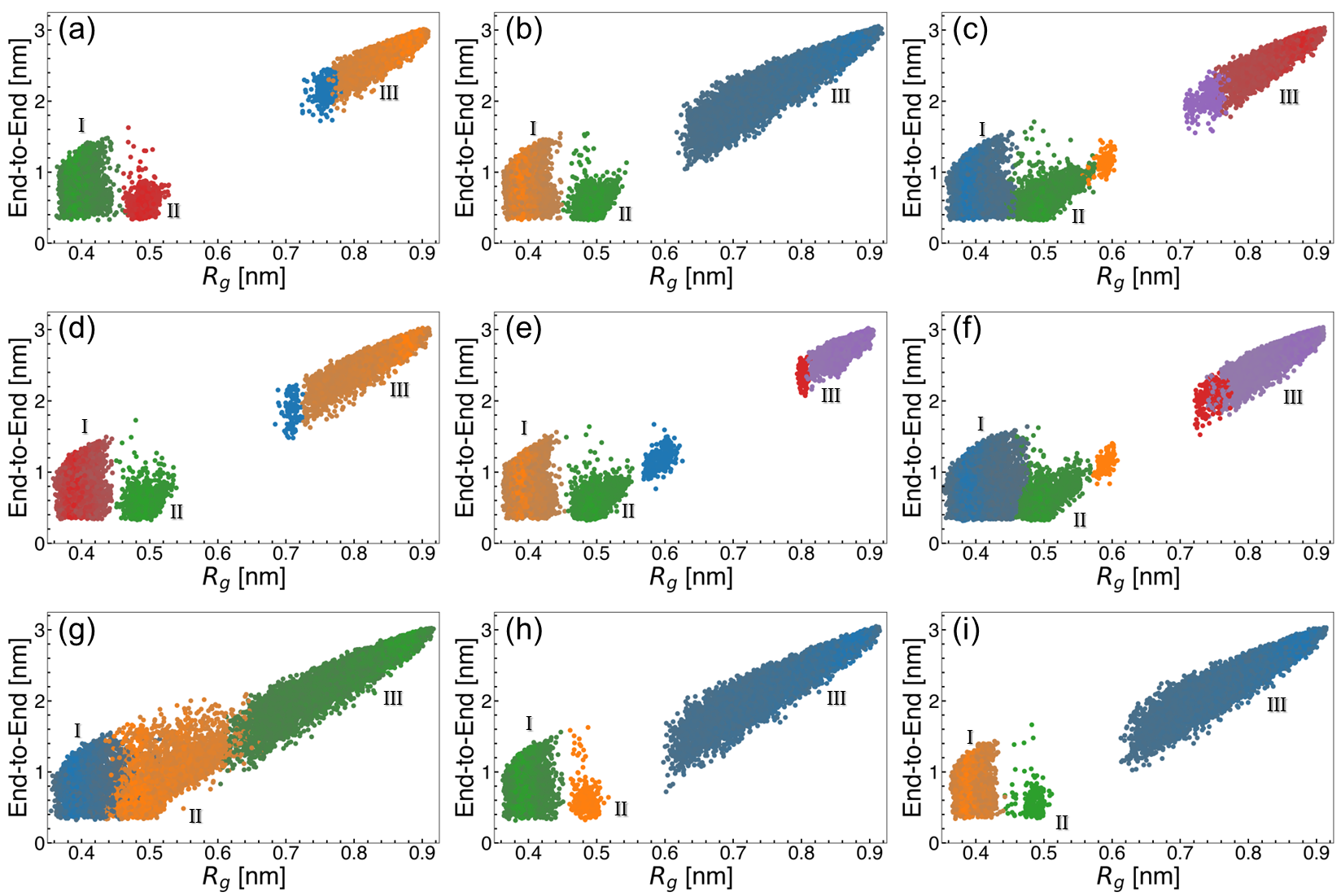}
   \caption{HDBSCAN clustering of polymer configurations for (a-c) Arg/Glu, (d-f) Arg/Lys, and (g-i) Lys/Glu solutions.} 
    \label{fig:si-HDBSCAN2}
\end{figure*}

\begin{scheme*}[!ht]
\centering
   \includegraphics[width=1.0\textwidth]{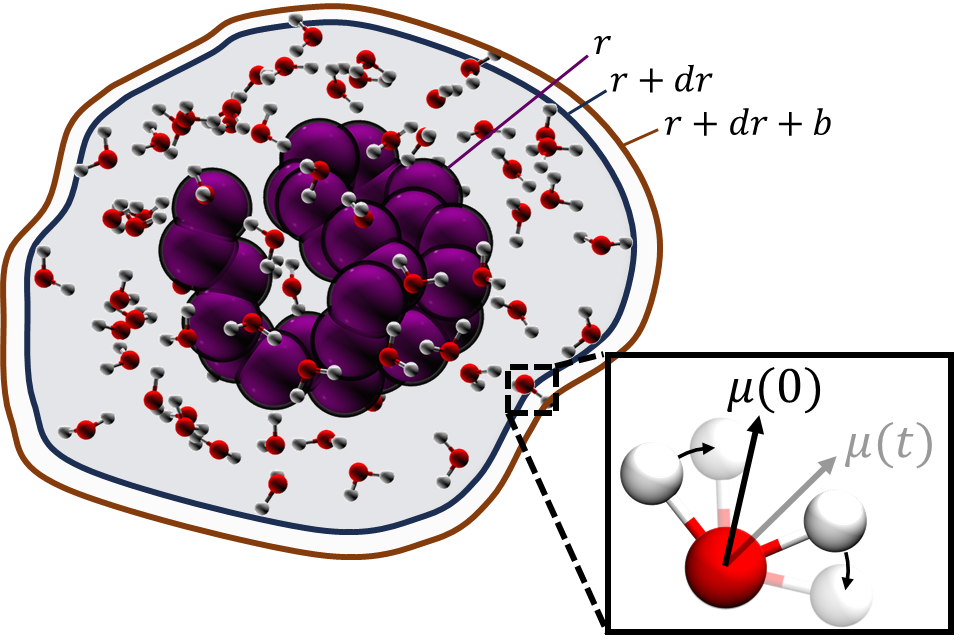}
   \caption{Schematic for water reorientation time calculation. The hydrophobic polymer is shown as purple spheres. $r$ denotes the van der Waals surface of the hydrophobic polymer, $r + dr$ denotes the region included in the calculation, and $r + dr + b$ denotes the furthest edge of the buffer region.} 
    \label{sch:si-watDyn}
\end{scheme*}

\begin{figure*}[!ht]
\centering
   \includegraphics[width=0.75\textwidth]{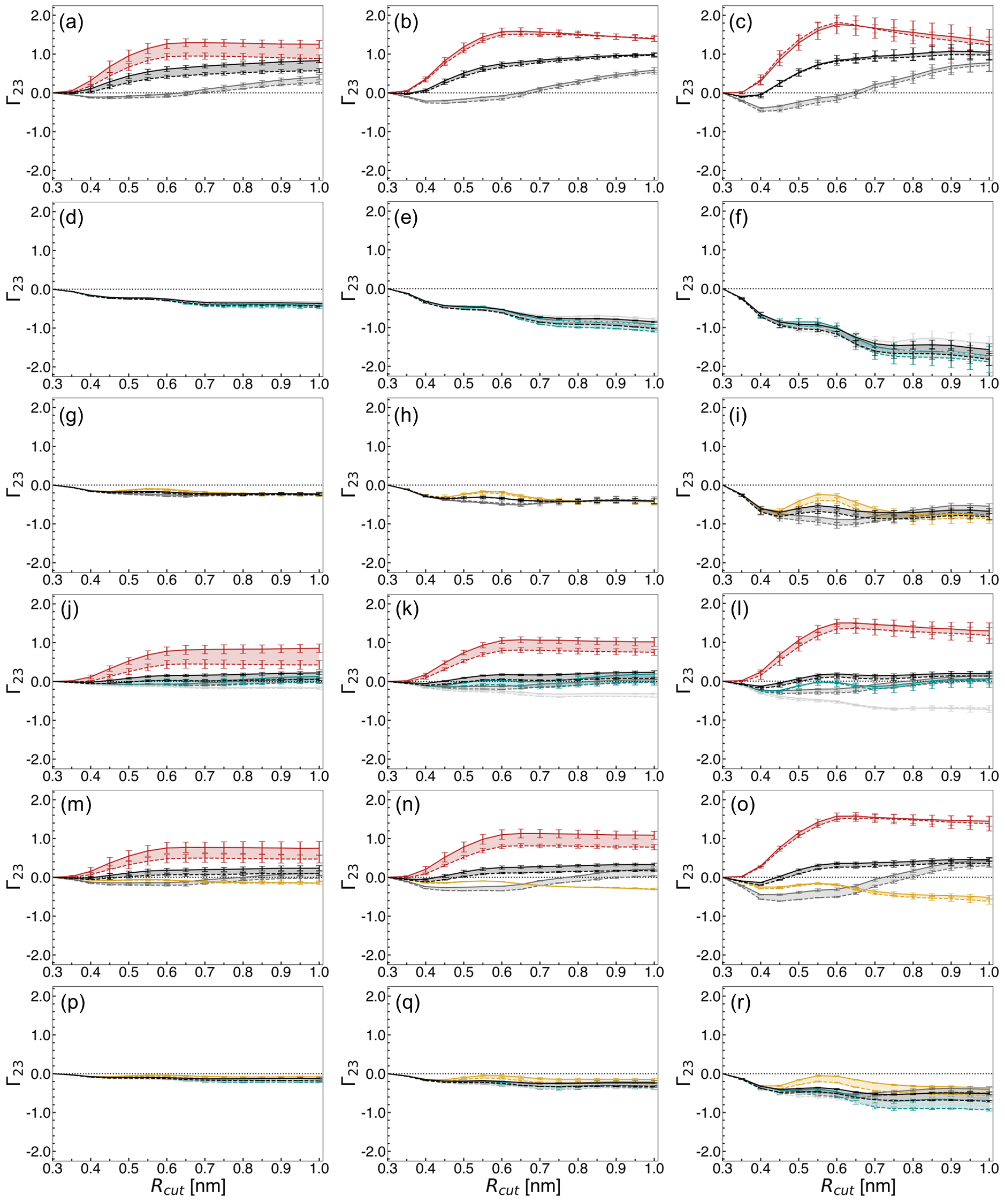}
   \caption{Preferential interaction coefficients for (a-c) Arg, (d-f) Glu, (g-i) lysine, (j-l) Arg/Glu, (m-o) Arg/Lys, and (p-r) Lys/Glu solutions. Arginine is colored in red, glutamate in blue, lysine in yellow, sodium in light gray, and chloride in dark gray. In all plots, the net preferential interaction coefficient is colored in black. Dashed lines indicate values for the unfolded state, while solid lines denote the folded state. Concentration increases from left to right in the order 0.25 M, 0.5 M, and 1.0 M. Mean values are reported from three replicate REUS simulations. Error bars were estimated as standard deviations from three replicate simulations.} 
    \label{fig:si-prefint}
\end{figure*}

\begin{figure*}[!ht]
\centering
   \includegraphics[width=1.0\textwidth]{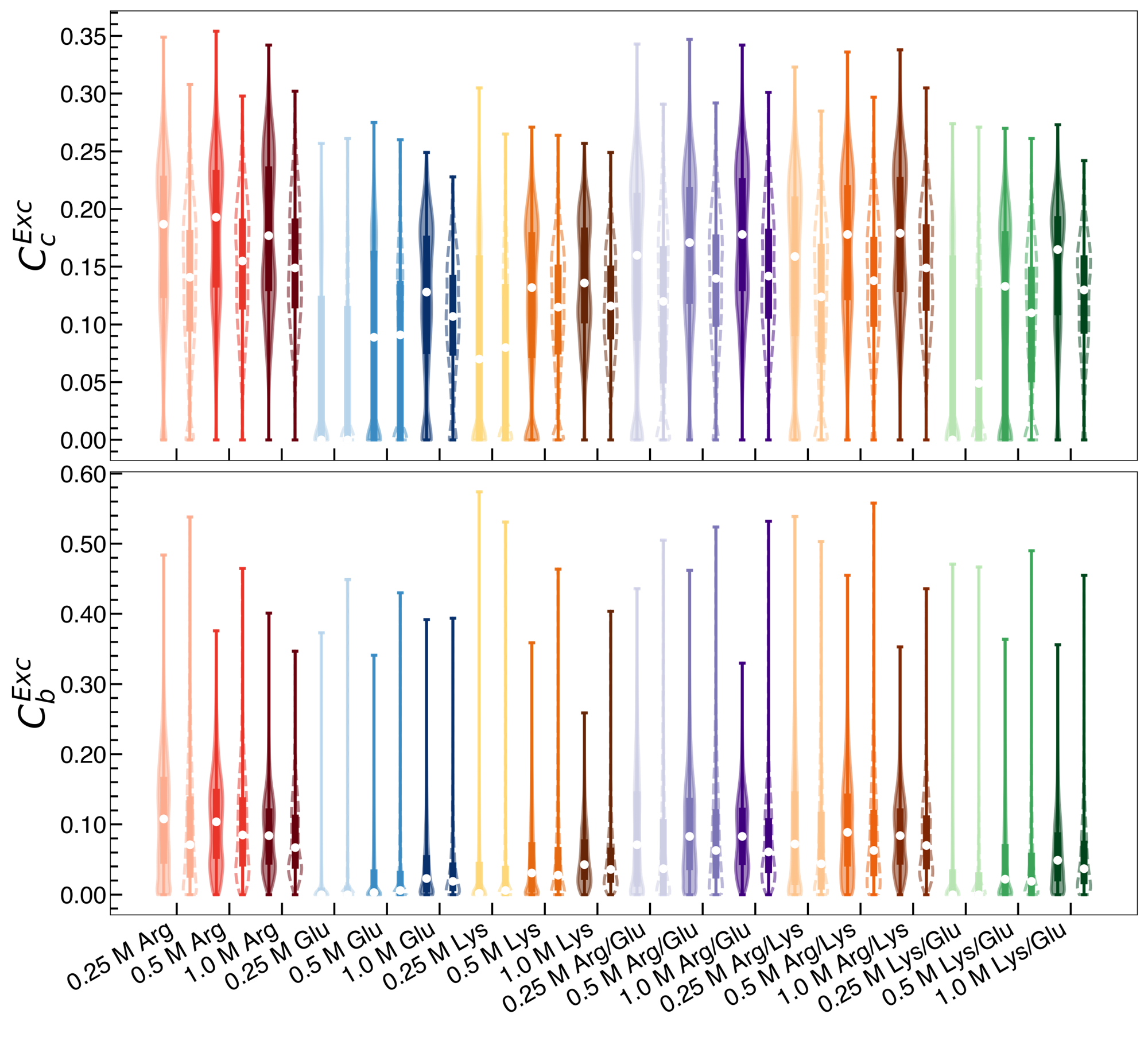}
   \caption{Excipient (top) closeness centrality and (bottom) betweenness centrality. Distributions with solid lines denote the folded ensemble, while dashed lines denote the unfolded ensemble.} 
    \label{fig:si-excCentrality}
\end{figure*}

\begin{figure*}[!ht]
\centering
   \includegraphics[width=1.0\textwidth]{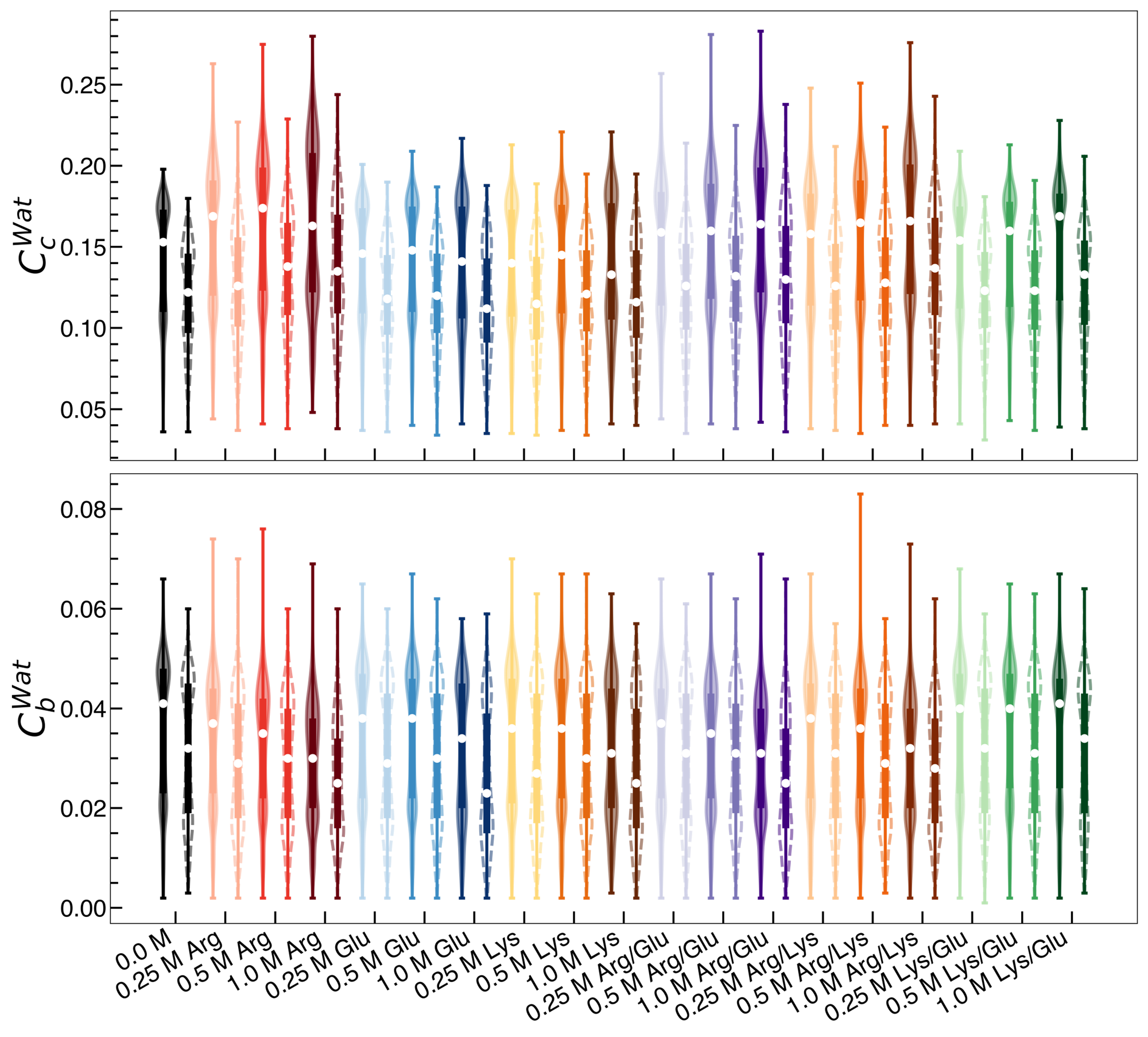}
   \caption{Water (top) closeness centrality and (bottom) betweenness centrality. Distributions with solid lines denote the folded ensemble, while dashed lines denote the unfolded ensemble.} 
    \label{fig:si-watCentrality}
\end{figure*}